\documentclass[12pt, draftclsnofoot, onecolumn]{IEEEtran}
\IEEEoverridecommandlockouts
\usepackage{multirow}
\usepackage{tabularx}
\usepackage{tikz}
\usepackage{url}
\usepackage{float}
\usepackage{array}
\usepackage{setspace}
\usepackage{hyperref}
\newcolumntype{P}[1]{>{\centering\arraybackslash}p{#1}}
\newcolumntype{M}[1]{>{\centering\arraybackslash}m{#1}}
\newcolumntype{B}[1]{>{\centering\arraybackslash}b{#1}}
\usepackage{amsmath,amssymb,amsfonts}
\usepackage[linesnumbered,ruled,vlined,commentsnumbered]{algorithm2e}
\usepackage{graphicx}
\usepackage{epstopdf}
\usepackage{textcomp}
\usepackage{gensymb}
\usepackage{hhline}
\usepackage{tikz}
\usepackage{xcolor}
\usepackage{cite}
\usepackage[utf8]{inputenc}
\usepackage{dirtytalk}
\usepackage[english]{babel}
\usepackage{colortbl}
\definecolor{LightGray}{gray}{0.85}
\definecolor{DeepGray}{gray}{0.25}
\usepackage{amsthm}

\theoremstyle{definition}

\theoremstyle{remark}

\theoremstyle{plain}
\newtheorem{assumption}{Assumption}

\usepackage{footmisc}

\usepackage{dsfont}
\usepackage[separate-uncertainty = true,multi-part-units = repeat, load-configurations = abbreviations, binary-units=true]{siunitx}
\usepackage{siunitx}
\usepackage{xcolor}

\usepackage{xcolor}

\usepackage{soul}

\usepackage[compact]{titlesec}
\IEEEaftertitletext{\vspace{-1.5\baselineskip}}

\begin{document}

\title{Scalable Multi-agent Reinforcement Learning Algorithm for Wireless Networks}
\author{
\IEEEauthorblockN{Fenghe Hu, Yansha Deng, and A. Hamid Aghvami}

\thanks{F. Hu, Y. Deng, and A. H. Aghvami are with  King's College London, UK (E-mail:{fenghe.hu, hamid.aghvami@kcl.ac.uk}).}
}
\maketitle

\begin{abstract}
    Reinforcement learning (RL) is known as a model-free and highly efficient intelligent algorithm and proved to be useful in solving radio resource management problems in wireless networks. However, for large-scale networks with high latency connection to center-server or capacity-limited backbone, it is not realistic to employ a centralized RL algorithm to perform joint real-time decision making for the entire network. The dimensional of the problem increases exponentially which introduces the scalability issue. Multi-agent RL, which allows separate execution of decision policy in each agent, has been applied to solve the scalability problem. In this paper, we propose a federated multi-agent RL 
    architecture for large-scale wireless scenarios, where access points (agents) share parameters to form consistency, save backbone traffic, and improve the convergence performance. Our results show that the federated frequency, which is critical for backbone traffic, has a limited effect on the convergence performance of the algorithm via an informational multi-agent analysis model. We also propose a transfer-learning based method to reduce the impact of biased local experience on the convergence performance during the federated process. We verify the analytical results of our proposed learning architecture via a simulated coordinate multipoint (CoMP) scenario.
\end{abstract}

\section{Introduction}
As a proposing research direction in future wireless networks, the cognitive or self-organised network should automatically make its decisions (radio resource management, routing, edge computing resource management, or etc) based on diverse service requirements via intelligent algorithms such as reinforcement learning (RL). One common approach is centralized RL, which employs a centralized controller to make all allocation or routing decisions. Although RL algorithms are model-free and capable to learn from environmental experience. The centralized RL requires to access information throughout the network to make the optimal decision. Unfortunately, when being applied to a large-scale network, such a centralized structure can cause significant overhead in backbone transmission and computation complexity, which makes it inapplicable \cite{Mismar2016Machine, AlEryani2021Multiple}. That is the so-called high-dimensional or scalability problem, which limits the use of RL in the large-scale wireless network. Existing solutions dealing with the large-scale network for the aforementioned scalability problem mainly uses fixed, greedy or game-theory algorithms, which are not sensitive to the network scale and usually leads to sub-optimal solutions \cite{Bassoy2017Coordinated}. 

To fully realise the benefit of RL in a large-scale network, researchers have tried to distribute the action decisions to entities inside the network, i.e. multi-agent RL, which are recognised as the key technology for the large-scale network. Such new learning algorithms allow each network entity (e.g. access-points (APs)) to make its own decisions distributively, while still optimising the common global target. However, unlike single-agent RL that has successfully solved many real-world problems with decent performance, multi-agent RL still suffers from convergence and instability problems due to the non-stationary environment. The non-stationary environment is caused by unknown information from other network entities, i.e. agents. Thus, directly extending the original centralized RL to multi-agent structures by simply allowing agents to make their decisions based on local observation usually brings limited performance \cite{Peng2021Multi}. Opponent modelling is an intuitive method to improve performance.
But the agents have to guess the possible action of their opponents, verse versa. This can create an infinitive logic guessing loop without ending \cite{He2016Opponent}. Another approach is to train a critic network with full knowledge of the environment conditioned on all agent's decisions, which is then used to guide the training of actor networks distributively in each agent \cite{Lowe2017Multi}. This approach only allows for the decision making to be performed at agents locally. It has theoretical proof but requires high backbone traffic and cannot fully solve the scalability problem. Another feasible solution is to acquire system-level information and train the RL for agents conditioned on local observations and executed distributively among agents, namely, centralized-training-distributive-execution \cite{Nasir2019Multi}. However, this approach is still limited by the connectivity between agents, for example, backhaul capacity.

Multi-agent RL is shown to be hard to design for scenarios where each agent can influence the global state. However, in the communication area, each user or base station has a limited coverage area, since the signal fades with the increasing of distance \cite{Yang2018Mean}, and the delay increases with the travelling distance and hops. This naturally limits the impact range of certain agents by its constrained coverage range and connected neighbours. Thus, agents can be seen as networked connected to their neighbours based on their impact range and strength. With this assumption, a supervised centralized graph neural network RL architecture has been proposed in \cite{Shen2021Graph} to capture such connections between agents, which verifies the effectiveness of the concept. However, it is hard to quantify the strength and topology structure of such a connection, especially in wireless cases. Another idea of mean-field theorem tries to use mean value to describe the impact of neighbourhoods \cite{Shiri2020Communication}. Following the idea of these works, we propose to apply convolutional layers to quantify the relationship between agents, as convolutional neural network (CNN) are shown effective in the signal and interference estimation in wireless communication \cite{Cui2019}.  

In the wireless communication area, although several works start using the multi-agent structure in wireless communication scenarios \cite{AlEryani2021Multiple, Peng2021Multi, Nasir2019Multi}. The theoretical analysis of the aforementioned process is missing, including gradient expressions, convergence proof, and convergence speed. They also lack detailed explanations of performance gain when introducing centralized training or federated learning into the algorithm. The convergence speed analysis is hard for neural network applications, especially for multi-agent RL. Luckily, the introduction of the informational model of multi-agent RL provides theoretical tools for convergence analysis \cite{Terry2020Revisiting}.

% But the centralized training or parameter sharing among all agents seems to have a significant benefit to the performance of multi-agent RL algorithms in some works \cite{Peng2021Multi}. Besides, the privacy requirement, which is common in the communication area, also motivate the sharing of parameters between agents instead of data or decision.

% Besides the problem mentioned previously, we also notice that nearly all RL in the communication area are episodic, which allows the reset of the environment or with a limited horizon. However, the interaction is a continuous procedure in the real world, which means no reset is possible and the horizon is infinity. This setting is recognized as non-episodic, which is critical for continuous control in real-time, for example, robot arm. The optimization target changes from the maximization of accumulative reward to average reward in this case, which is shown significant performance difference in the performance of the RL algorithm \cite{CoReyes2020Ecological, Naik2019Discounted}. There are limited discussions in this area.

Inspired by the aforementioned ideas, we integrate the ideas of centralized-learning-decentralized-execution and networked communication entities in our work. Instead of training the model in a centralized entity, we apply federated learning to share knowledge among agents while keeping the training locally by sharing parameters. 
In this paper, we introduce a federated multi-agent RL architecture to address the scalability problem with convolutional layers. The agents train their policy reference on their impact range (signal coverage) distributively while sharing the parameters of their models to accelerate the learning.
% We illustrate the design principle of the proposed architecture with the characteristic of the communication problem. We also give the convergence proof of federated multi-agent algorithm and try to explain the reason why federated learning can support the performance of multi-agent RL. Besides, we also derive an upper bound of converging with the informational model for the multi-agent algorithm, which shows the influence of federated frequency on the converge performance. Subsequently, we exam our architecture and theoretical conclusions in a simulated coordinated multi-point (CoMP) environment, which is a typical case for cooperation techniques in the cellular network. According to the simulation results, we also show the influence of the non-episodic environment on current RL algorithms and our architecture can effectively handle the non-episodic case. 
The contributions of this paper are summed as follows:
\begin{itemize}
    \item We propose a federated distributive multi-agent RL architecture to solve scalability problems in large-scale wireless scenarios and perform the theoretical analysis and convergence proof to show the benefit of integrating the federated into a multi-agent system.
    \item We derive the upper bound of convergence speed for our proposed federated multi-agent RL with the informational model, which is critical for balancing the trade-off between convergence performance and backbone traffic in federated learning.
    \item We highlight the problem of centralized-decentralized-mismatch \cite{Wang2020Policy} in the multi-agent environment and analyse the impact of this phenomenon in our introduced federated multi-agent RL. To reduce its negative impact, we also propose a transfer-learning based federated method.
    \item We implement our proposed federated multi-agent RL architecture with different state-of-the-art RL algorithms in a simulated coordinated multi-point (CoMP) scenario \cite{CoReyes2020Ecological, Naik2019Discounted}. With the actor-critic algorithm, we show that our proposed multi-agent architecture can effectively optimise the cooperation decisions in the CoMP scenario, and the federated learning can support and improve the learning performance. Especially, transfer learning is shown to significantly reduce the impact of centralized-decentralized-mismatch. We also verify our analysis results for the impact of federated frequency on the convergence speed via simulation. 
    % \item Highlight the necessity of investigating the performance difference between episodic and non-episodic learning, as the environment can't simply reset or restart. 
\end{itemize}

\section{A Coordinated Multipoint Wireless Networks Model}
In this section, to ease the understanding of the basis of our algorithm, we first present the system model and basic information of a radio resource management optimization problem. As a specific example, we consider the joint-transmission coordinated multipoint (CoMP) in a large-scale multi-cell network, which faces scalability problems when applying RL methods.

\subsection{System Model}
We consider the joint-transmission CoMP (JT-CoMP) for downlink transmission with a set of access-points (APs), denoted by $\mathcal{B}$. For simplicity, each AP is located in hexagonal-grid and equipped with one omnidirectional antenna for downlink transmission. All APs are connected via fibre links, which allow data sharing through a central unit. A set of users, denoted by $\mathcal{U}$, are located in the serving area following the Poisson point process (PPP). As shown in Fig. \ref{fig:scenario}, to enhance the quality-of-service (QoS) for the cell-edge users, the neighbouring APs seek to form cooperative clusters \cite{Geo2019Coordination}, where the signals are transmitted and enhanced by cooperative APs using the same frequency band. 
Through joint transmission, the CoMP technology enhances the cell-edge users' QoS at the cost of backhaul overhead and frequency resource \cite{Shami2018User}. The larger the cluster size, the more effective cooperation among APs, and the higher backhaul capacity and frequency resource requirements. In particular, the users benefit from a large cluster and effective collaboration among APs. However, with limited AP capability and backhaul capacity, the number of cooperative APs is also limited. Also, the collaboration requires reserving frequency band for cell-edge users in both APs \cite{Shami2018User}. Due to this trade-off, it is common to have a maximum cluster size of $B_{\max}$ \cite{Guidolin2014Distributed}.

\begin{figure}[!htbp]
    \centering
    \includegraphics[width=0.48\textwidth]{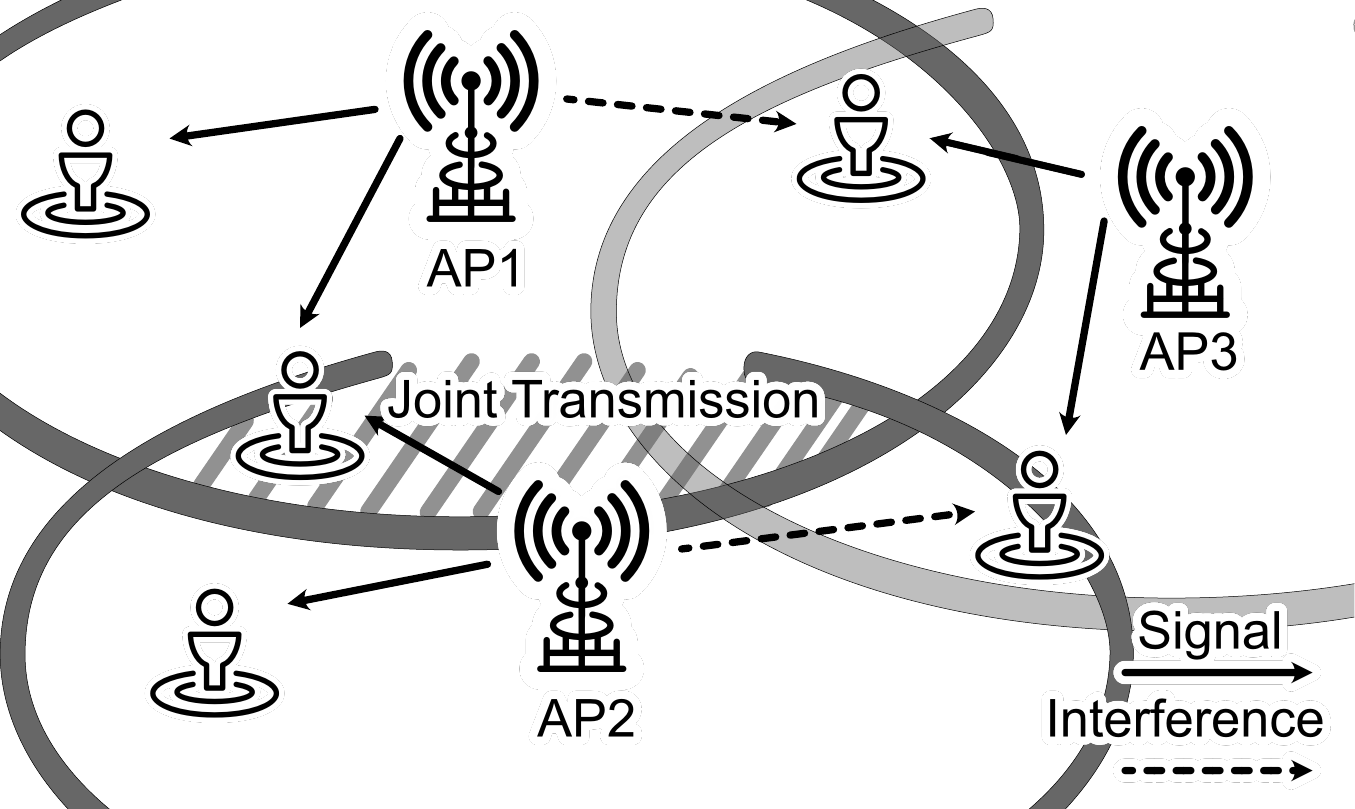}
    \caption{JT-CoMP Scenario with AP1 and AP2 forming a joint transmission group to enhance the service for users in overlapped effective area while causing interference to users serving by AP3.}
    \label{fig:scenario}
\end{figure}

Serving by such network, we consider that a set of users requests a certain amount of data $D$ from APs in the downlink. The request is considered to be failed if it is not satisfied within a certain period $T$.
Then, we quantify the service performance of the network with a certain cooperation policy $\pi$ in $t$ time slot via a QoS function, which is usually a function of resulting users' signal-to-interference-noise-ratio and requested data amount $D$. One common QoS function in radio resource management problems is a spectrum efficiency function with the consideration of service outage \cite{Chen2020Federated}, which is used in the simulation part:
\begin{equation}
    {r}^u_t=
    \begin{cases}
        \min\{\log_2(1+\underbrace{\frac{\sum_{i\in \mathcal{B}^u}P_i\beta_{i,u}d_{i,u}^{-\alpha}}{\sum_{i\in \mathcal{B}^{\mathcal{U}/u}}P_i\beta_{i,u}d_{i,u}^{-\alpha}+\sigma^2}}_{\text{SINR}_u}), D\}, t^u \leq T \\
        0, t^u > T
    \end{cases}
    \label{qos}
\end{equation}
where $P_i$ is the transmit signal power from $i$-th agent, $\beta_{i,u}$ is the small-scale fading factor, $d_{i,u}^{-\alpha}$ denotes the large-scale fading that depends on the distance and path loss factor $\alpha$, $\mathcal{B}^u$ is the set of associated APs (include the cooperation APs) of user $u$, $\mathcal{B}^{\mathcal{U}/u}$ is the set of APs which are not associated to user $u$, $\sigma^2$ is the noise power. The user will be removed from serving area when its waiting time $t^u$ exceed maximum tolerable time $T$ or successfully receives $D$ data.

For the cooperative CoMP problem, we define a policy $\pi$ to decide which APs should belong to the same cluster dynamically over time. In this way, the optimization problem of the CoMP clustering scenario can be written as
\begin{equation}
\begin{aligned}
    \max\limits_{\pi} &\sum_t^\infty[\sum_{u\in\mathcal{U}}r^u_t|\{\mathcal{B}^u\}_{u\in\mathcal{U}}\sim\pi],  s.t. \quad |\mathcal{B}^u|\leq B_{\max}\\
\end{aligned}
    \label{qos_opt}
\end{equation}
where $r^u_t$ is defined in Eq.\eqref{qos}, $\mathcal{B}^u$ denotes the set of APs serving $u$th user cooperatively, which is decided via policy $\pi$, and $B_{\max}$ is the maximum cluster size, which is usually around $3$ \cite{Geo2019Coordination}. 

For our considered optimization problem, there are commonly two existing approaches:
\begin{itemize}
    \item The greedy algorithm is the most widely used approach due to its low complexity and easy implementation. By selecting the AP with the maximum CoMP gains to cooperate, the cooperation decisions are made greedily one by one from a randomly chosen AP and propagate through the whole network, thus the later formed clusters may be sub-optimum due to the lack of global consideration.
    \item The game theory approach applies simple merge-and-split rules among APs distributively, which significantly reduce the signal overhead \cite{Guidolin2014Distributed}. However, the complexity also increases with the number of cooperative APs, and its performance heavily depends on the precise estimation of the QoS value.
\end{itemize} 
We highlight that both of the aforementioned algorithms aim to design a simple policy and apply them to each AP distributively, due to that the size of the cooperation problem increases with the number of cooperative APs. The problem is shown to be NP-hard \cite{Solaija2020Generalized}, and such types of scenarios are common in large-scale networks, while more cooperation between APs is a foreseeable trend. 

\subsection{Problem Decomposition via Impact Range}
To address the aforementioned limitations, deep reinforcement learning (DRL) with the neural network, demonstrate its capability to manage the cooperation among APs with the knowledge learnt from the environment. Most existing DRL algorithms solve cooperation problems by making dynamic cooperation decisions via a centralized controller, but its poor scaling capability, high computation complexity, and significant communication and computation latency between controller and APs prevent the use for the large-scale multi-cell scenario. Luckily, multi-agent RL may address the high dimensional decision-making problem by allowing each entity (AP) to optimize its long-term performance by interacting with other agents and environment  \cite{Zhang2019Decentralized}. Inspired by the idea of solving the original large-scale problem distributively, we identify several key properties of problem decomposition in a large-scale multi-cell network in the following, which can reduce the complexity and motivate new design and analysis of multi-agent RL algorithms.

We first note that the QoS function in Eq. \eqref{qos} is geometrically separable and independent. The QoS value only depends on the states of local users (the SINR of user $u$), which is affected by the cooperation decisions and the signal strength received at that location. In this way, the QoS value in different locations is independent of each other. 
Second, the wireless signals fade with the increasing distance or the existence of variant obstacles. Each AP imposes limited signal gain/interference to the surrounding area. Thus, each AP has an effective region, that is limited by its maximum coverage area. The QoS value of the effective region precisely reflects the performance of local AP's and neighbours' policies. We define the effective region of $i$th AP as the set of users inside the $i$th AP's effective region, denoted as $\mathcal{U}^i$:
\begin{equation}
    \mathcal{U}^i = \{u|P_id_{i,u}^{-\alpha} \geq \sigma\}\quad \forall u\in\mathcal{U},
\end{equation}
where $\sigma$ is the average receive power threshold of the considering $u$th user is in $i$th AP's effective region \cite{Shami2018User}, and we have $\mathcal{U}^i\subset \mathcal{U}$. For simplicity, the effective region can be considered as a circle because of line-of-sight transmission. 

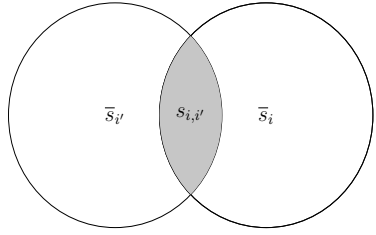
\begin{figure}[t]
    \centering
    \resizebox{0.3\textwidth}{3cm}{
    \begin{tikzpicture}
        \draw (0,0) circle (2.12cm);
        \draw (3,0) circle (2.12cm);
        % \draw (6,0) circle (2.12cm);
        % \draw (3,3) circle (2.12cm);
        % \draw (3,-3) circle (2.12cm);
        \draw (0,0) node {$\overline{s}_{i'}$};
        \draw (3,0) node {$\overline{s}_i$};
        % \draw (6,0) node {$\overline{s}^{i^*}$};
        % \draw (3,3) node {$\overline{s}^{i''}$};
        % \draw (3,-3) node {$\overline{s}^{i'''}$};
        \draw [clip](3,0) circle (2.12cm);
        \fill[gray!45] (0,0) circle (2.12cm);
        % \fill[gray!45] (6,0) circle (2.12cm);
        % \fill[gray!45] (3,3) circle (2.12cm);
        % \fill[gray!45] (3,-3) circle (2.12cm);
        \draw (1.5,0) node {$s_{i,i'}$};
        % \draw (4.5,0) node {$s^{i,i^*}$};
        % \draw (3,1.5) node {$s^{i,i''}$};
        % \draw (3,-1.5) node {$s^{i,i'''}$};
    \end{tikzpicture}}
    \caption{The relationship between overlapped effective region $s_{i,i'}$ and $\hat{s}_i$ for two neighboring APs.}
    \label{fig:relation}
\end{figure}

Within each AP's effective region, we can further decompose the local QoS function into one region which is majorly affected by the AP itself, and another region that is jointly affected by the neighbouring APs with the overlapped effective regions. This is because the distant APs cause little gain/interference. We plot the relationship between APs and their overlapped effective region as Fig.\ref{fig:relation}. Here, for notation simplicity, we only show the case, where the overlapped region is affected by at most two APs. For a set of AP $\mathcal{B}=\{i,i'\}$, the state of $i$th AP's effective region is split into the regions with and without overlapping with neighboring $i'$th AP's effective region, denoted as $\overline{s}_i$ and $s_{i,i'}$, respectively ($s_i=\overline{s}_i\cup s_{i,i'}$).  
Then, we can write the local QoS function of $i$-th AP with the users inside its effective region as
\begin{equation}
\begin{aligned}
    & r^i_t = r^i(\overline{s}_i, a_i) + \sum_{i'\in \mathcal{B}^{-i}}r^i(s_{i,i'}, a_i, a_{i'}) = \sum_{u\in\mathcal{U}^i}r^u_t,
\end{aligned}
\label{reward_sep}
\end{equation}
where $\mathcal{B}^{-i}$ denotes the set of APs with overlapped effective region, for Fig. \ref{fig:relation} $\mathcal{B}^{-i}=\{i'\}$, $r^i_t$ is the local QoS function of the AP $b$ in $\mathcal{B}$, $\overline{s}_i$ is the state information near the AP $b$ without overlapping effective region, $s_{i,i'}$ presents the state of users in overlapped effective region, whose QoS is affected by APs from both sides. Then it is clear to see that $r^i$ contains two parts, one refers to the dominated status $\overline{s}_i$ and another refers to the overlapped status $s_{i,i'}$.

% Based on aforementioned properties, it is possible to model the effects between APs in considering communication network as a graph $<\mathcal{B},\mathcal{G}_t>$ with a set of APs (vertex) $\mathcal{B}$, and a set of directional weighted edge set $\mathcal{G}_t$ at time $t$. The weight of each edge presents the signal gain/interference to users in overlapping effective regions. Then, our considered optimization problem can be described as a canonical characteristic from the coalitional game, where the gain of cooperation only correlates to the participant members. It should be noted that the graph is sparse due to the limited coverage range of each AP. Besides, it also has been shown that the graph has the property of permutation invariance \cite{https://ieeexplore.ieee.org/stamp/stamp.jsp?tp=&arnumber=9252917}.
With the aforementioned properties, our considered optimization problem can be decomposed into identical sub-problems from the view of individual APs, which optimises their local QoS function by interacting with their opponents and the environment. This matches the idea of value decomposition network \cite{Rashid2018QMIX}. By doing so, the complexity for each distributive policy in each AP is largely reduced, and the decisions can be generated directly without communicating with centralized processors. This forms the basic idea to deal with the high dimensional large-scale communication environment.

\section{Multi-agent Reinforcement Learning Design with Problem Decomposition}
In this section, we aim to develop a scalable multi-agent RL architecture for large-scale communication problems based on the aforementioned decomposition method.

\subsection{Stochastic Game Definition}
To solve our considered problem with RL methods, we first define our problem as a networked stochastic game, which can be characterized by a tuple of $<\mathcal{S},\mathcal{B},\{\mathcal{O}^i\},P,\{\mathcal{A}\},\{\mathcal{A}^i\},\{\mathcal{R}^i\},\Omega>$. We define each component of this tuple notations as
\begin{itemize}
    \item $\mathcal{S}$ is a set of joint state ($s\in\mathcal{S}$), and $\mathcal{S}^i$ represents the set of local state of agent $i$ ($s^i\in\mathcal{S}^i$). $S_t$ is the state at time $t$, which includes users' position, SINR, neighboring AP cooperation state, and AP's transmit power, etc.
    \item $\mathcal{B}$ is the set of agents ($b\in\mathcal{B}$), which co-located with each AP,
    \item $\mathcal{O}^i$ is a set of local observations of the $i$-th agent ($o^i\in\mathcal{O}^i$), which contains users' location, neighboring AP's location and request status in our considered CoMP scenario,
    \item $P$ is a transition probability function which maps the state-action to the next state, i.e. $P(s'|s,a):\mathcal{S}\times\mathcal{A}\times\mathcal{S}$,
    \item $\Omega$ is the observation function, which maps the local state of agent $i$ to its observation, i.e. $\Omega(o^i|s_i):\mathcal{S}^i\times\Omega\rightarrow [0,1]$, that is decided by APs' sensors' capability. Due to the limited sensor capability, the observation at the AP usually contains less information than the original state. In this way, the problem can be analysed as a fully-observable problem with $\Omega(o^i|s_i)$ added to policy and transition probability.
    \item $\mathcal{A}$ is the set of joint actions of agents ($a\in\mathcal{A}$) of all APs. The action $a^i$ of $i$th AP is the cooperation decision, which reflects its request to cooperate with a certain number of the neighbouring APs. By exchanging the requests among APs, a cooperation cluster is formed when both APs achieve an agreement in cooperation. The local action set of the $i$th AP is given as $\mathcal{A}^i$ ($a^i\in\mathcal{A}^i$). In our considered scenario, the size of local action space is defined as $|\mathcal{A}^i|=\sum_{c=0}^C|\mathcal{B}^{-i}|!/(c!(|\mathcal{B}^{-i}|-c)!)$, where $C$ is the maximum cluster size. It is worth mentioning that the joint action space increases exponentially with the number of cooperative APs, i.e $|\mathcal{A}|=(|\mathcal{A}^i|^{|\mathcal{B}|})$, which results in the scalability problem.
    \item The QoS $r^i$ (given in Eq.\eqref{reward_sep}) is used as the reward function for the $i$th agent, i.e. $r^i(s,a):\mathcal{S}\times\mathcal{A}\rightarrow \mathds{R}$. $r$ is the overall sum reward for all agents.
\end{itemize}

In each round of decision, each AP observes its surrounding environment in state $S_t$. Then, each AP chooses its action according to its local policy $\pi_{\theta^i}$ defined by a set of parameters $\theta^i$, where the probability of choosing action $a^i$ with observation $o^i$ is represented as $\pi_{\theta^i}:\mathcal{O}^i\times\mathcal{A}^i\rightarrow[0,1]$. Then, the cooperation decision $A_t$ is formed with actions from all APs. Usually, in a centralized approach, such cooperation decision (action) is made by a central utility based on information collected from all APs. This process consumes at least a round-trip delay from AP to central utility, which is not ideal. To fully explore the benefit of distributed algorithms, we design a hand-shake mechanism for our scalable multi-agent architecture to enforce the maximum cluster size constrain in Eq.\eqref{qos_opt}. The hand-shake mechanism allows agents to fully control their cooperation, which ensures the constrain and removes the information exchange between agents and the central server. To form the cooperation cluster, the agents in the cluster are required to agree on the cooperation via hand-shake. The cooperation cluster is formed when both neighbouring agents send the cooperation request (action) to each other. By performing the hand-shake, the cooperation decision for the whole network can be made via one single round-trip information exchange between neighbouring agents. This significantly enhances the scalability of our design, but the hand-shake introduces challenges in algorithm design.

\begin{figure}[!htbp]
    \centering
    \includegraphics[width=0.48\textwidth]{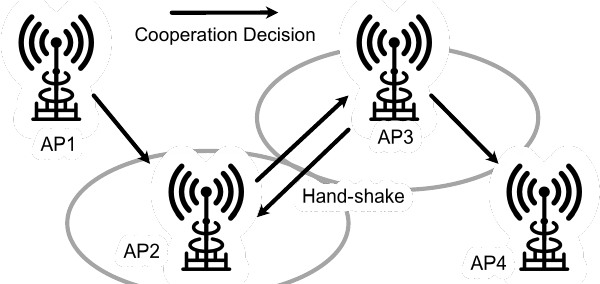}
    \caption{The handshake process for cluster forming. The AP2 and AP3 form cooperation cluster by sending requests to each other, while AP1 and AP4 fail to form cluster without the agreement from the AP2 and AP3.}
    \label{fig:scenario}
\end{figure}

After the users are served under current cooperation decisions for a certain time slot. Each AP observes a local reward, which measures the efficiency of the current cooperation decision. The reward is then used to update the local policy.
Then, the considered system shifts to a new state $S_{t+1}$, while all APs observe their observation $O_{t+1}^i$ from state $S_{t+1}^i$ based on function $\Omega$. Then the aforementioned process is repeated and the state-action pair forms a trajectory $\tau$. 

% It is worth mentioning that the centralized action selection, which maintains a single policy and selects the cooperation action for all APs, faces the problem of dimension explosion, where the number of possible actions and the size of required parameters grows exponentially with the number of participating APs. It is nearly impossible to design and train a centralized policy in our considered scenario. 
To conclude the properties mentioned in the aforementioned stochastic games and simplify further discussions, we make the assumption on the transition functions and policy functions of agents, which is reasonable and standard for neural network based RL algorithms.
\begin{assumption}
We assume that the actions selected by different agents are statistically independent. Thus, the joint policy $\pi$ of all agents is factorized as the product of all local policies, i.e. $\pi_\theta = \prod_{i\in\mathcal{B}}\pi_{\theta^i}(o^i,a^i)$. Also, the policy function is differentiable with respect to all possible parameter $\theta^i$. As such, we can write the state transition probability between two state $s$ and $s'$ ($s,s'\in\mathcal{S}$) under a joint policy $\theta$ as
\begin{equation}
    \mathds{P}_\theta(s'|s) = \sum_{a\in\mathcal{A}}\prod_{i\in\mathcal{B}}\pi_{\theta_i}(o_i,a_i)\Omega(o_i|s_i) P(s'|s,a),
\end{equation}
where $\theta=[\theta_i]_{i\in\mathcal{B}}$. The proposed Markov chain is irreducible and aperiodic under any policy set $\pi_\theta$. To simplify the notations, we write $\pi_\theta(s,a)=\prod_{i\in\mathcal{B}}\pi_{\theta_i}(o_i,a_i)\Omega(o_i|s_i)$, which introduces the partially observable cases into proposed stochastic game \cite{Shalabh2009Natural}.
\label{basic_ref_AC}
\end{assumption}
The Markov chain is irreducible and aperiodic means that it has a stationary distribution of the existence of state $s$ under the policy defined by $\theta$, which is denoted as $d_\theta(s)$ for any $s$. These assumptions are critical for methods like policy gradient and are satisfied by policies defined by neural network parameters. Then, we write our long-term optimization goal for the considered stochastic game by introducing joint policy $\pi_\theta$ and rewriting the QoS as a reward in Eq.\eqref{qos_opt}:
\begin{equation}
    \max\limits_{\theta} J(\theta)=\mathds{E}_{(s,a)\sim \mathds{P}_\theta(s, a)}[\sum_{t=0}^T r(s,a)], |\mathcal{B}^u|<B_\text{max}
\end{equation}
where $P_\theta(s, a) = $ is the probability with state-action pair $(s,a)$.

\subsection{Partial Derivation of Policy Gradient Method}
In this section, we apply the policy gradient methods to solve our considered problem and present the gradient update steps. The optimization target is to find a $\theta^*$ which maximizes the target function:
\begin{equation}
\begin{aligned}
    \theta^* & = {\arg\max}_\theta J(\theta)= {\arg\max}_\theta\mathds{E}_{(s,a)\sim \mathds{P}_\theta(s, a)}[\sum_{t=0}^Tr(s,a)]
\end{aligned}
\end{equation}

Then, the optimization algorithm updates $\theta$ in the direction of the gradient:
\begin{equation}
\begin{aligned}
    \nabla_\theta & J(\theta) = 
    \begin{cases}
        &\mathds{E}_{\tau\sim \mathds{P}_\theta(\tau)}[Q_\theta(S_t,A_t)\nabla_\theta \log \pi_\theta(A_t|S_t) | (S_t,A_t)\sim\tau], T\leq\infty\\
        &\mathds{E}_{(s,a)\sim \mathds{P}_\theta(s,a)}[Q_\theta(s,a)\nabla_\theta \log \pi_\theta(a|s)],  T=\infty.
    \end{cases}
\end{aligned}
\label{gradient_update}
\end{equation}
where $\mathds{P}_\theta(\tau)$ is the probability of trajectory and $\mathds{P}_\theta(\tau)=\mathds{P}(S_0)\prod^T_{t=1}\pi_\theta(A_t|S_t)\mathds{P}_\theta(S_{t+1}|S_t,A_t)$, and $Q_{(\tau,\theta)}(S_t,A_t)$ is called the state-action function (Q-function) for state-action pair $(S_t, A_t)$ in trajectory $\tau$, which counts for the sum of future reward from $(S_t,A_t)$ to the end of $\tau$, i.e. $Q_\theta(S_t,A_t) = \mathds{E}_{(S_{t'},A_{t'})\sim \mathds{P}_\theta(S_{t'},A_{t'})}[\sum_{t'=t}^T r(S_{t'},A_{t'})]$. Usually, a episodic case is considered where the environment can be reset after $T$ actions, i.e. $T<\infty$. In many real-world cases, such reset is impossible, i.e. $T=\infty$. In these non-epsodic cases, the Q-function is considered as the expected reward of overall possible state-action pair under the policy $\pi_\theta$.

% In this way, the update function can also be written as
% \begin{equation}
%     \nabla_\theta J(\theta) = \mathds{E}_{(s,a)\sim \mathds{P}_\theta(s,a)}[\underbrace{(Q_\theta(s,a))-V_\theta(s))}_{A_\theta(s,a)}\nabla_\theta \log \pi_\theta(a|s)] \text{ or } \mathds{E}_{(s,a)\sim \mathds{P}_\theta(s,a)}[(Q_\theta(s,a))-\overline{r})\nabla_\theta \log \pi_\theta(a|s)],
% \end{equation}
% where $A_\theta(s,a)$ is known as advantage function which presents how good state-action pair $(s,a)$ is. In this way, the parameter $\theta$ can be updated with the gradient and Q-function.

Inspired by the gradient update in Eq.\eqref{gradient_update}, it is possible to employ a centralized estimator to fit the Q-function precisely with global knowledge. However, such algorithms require frequent communication between the central server and agents. The way of the global Q-function estimation can be time-consuming and unrealistic in our considered large-scale network with hundreds of agents and super-wide serving area, i.e. scalability problem.

To train multiple agents simultaneously and distributively, we try to decompose the problem of updating $\theta$ parameters into local updates for each agent, while still solving the global optimization problem. We first formulate the local update for $\theta_i$ in each agent by taking partial derivation of original target function with respect to the local parameter $\theta_i$. Recall that Assumption \ref{basic_ref_AC} and the definition of observation function, the policies in each agents are independent. Then, we have
\begin{equation}
\begin{aligned}
            \nabla_{\theta_i}&\log\pi_{\theta}(a|s) \overset{(a)}{=} \nabla_{\theta_i} \log \pi_{\theta_i}(a_i|o_i),
\end{aligned}
\end{equation}
in $(a)$ we apply the definition of joint policy in the Assumption.\ref{basic_ref_AC}, and $\nabla_{\theta_i}\log\Omega(o_i|s_i)=0$ following the definition that observation function does not correlates with $\theta$. 

We then take the partial derivation of target function $J(\theta)$ in Eq. \eqref{partially_derived}, where $\mathds{P}_\theta(S_0\rightarrow s,t)$ is the probability of having state $S_t=s'$ from $S_0=s$ at time $t$ under the policy defined by parameter $\theta$, $d_\theta(s) = \sum_{t=0}^\infty \mathds{P}_\theta(S_t=s)$. We have $(a)$ by continually expanding the equation. The function can be simplified according to Assumption \ref{basic_ref_AC}.
\begin{figure*}[!htbp]
\begin{equation}
\label{partially_derived}
    \begin{aligned}
                \nabla_{\theta_i} J(\theta) 
                \overset{(a)}{=} 
                \begin{cases}
                    \sum_{t=0}^T \mathds{E}_{s'\sim \mathds{P}_\theta(S_0\rightarrow s,t)}[\mathds{E}_{a\sim \pi_\theta(a'|s')}[\nabla_{\theta_i} \log\pi_{\theta_i}(a_i'|o_i') Q_{\theta}(s',a')]], &T < \infty\\
                    \sum_{s\in\mathcal{S}} d_\theta(s)\sum_{a\in\mathcal{A}}\nabla_{\theta_i}\pi_{\theta_i}(a_i|s_i)Q_{\theta}(s,a) = \mathds{E}_{(s,a)\sim \mathds{P}_\theta(s,a)}[\nabla_{\theta_i} \log\pi_{\theta_i}(a_i|o_i) Q_{\theta}(s,a)]], &T=\infty
                \end{cases}
    \end{aligned}.
\end{equation}
\end{figure*}

The result of Eq.\eqref{partially_derived} shows that the update in local parameter sets $\theta_i$ can still optimise the overall optimization problem if we can obtain a correct estimation of the global Q-function $Q_{\theta}(s, a)$. 

\subsection{Decomposition in Q-function Estimating}
In the following, we propose the idea of Q-function decomposition considering several unique characteristics in wireless networks, which allow the distributive estimation of Q-function. We first recall that the agent has the limited effective region in the CoMP scenario. Thus, we can rewrite the global Q-function in Eq.\eqref{gradient_update} as the local Q-function
\begin{equation}
    Q_{\theta_i}(s,a) \approx Q_{\theta_i}(s_i, a_i, \{a_{i'}\}_{i'\in\mathcal{B}^{-i}}) 
\label{effective_limit}
\end{equation}
where $\mathcal{B}^{-i}$ is the set of neighbouring agents, which have overlapped coverage regions with agent $i$.
From the Eq.\eqref{effective_limit}, we can see it is possible to optimise the algorithm with known neighbouring actions by performing the gradient update for the policy with Q-function estimation, as the environment can be seen as stationary with known neighbours' actions. However, action sharing can be hard in certain cases, i.e. unmanned vehicle cooperation. The neighbouring agents with the overlapped effective region are hard to identify and can be outside of the communication range.

Without the information from neighbouring agents, we need to quantify the influences and estimate the neighbours possible actions. We first leverage the properties that the Q-function is geometrically in-correlated and independent (QoS function in Eq.\eqref{qos}) in typical wireless communication scenarios. Then, it is possible to decompose the centralized estimator and allow the distributed estimation in each agent. Combing the effective region for agents in the communication scenario mentioned in the previous section and Fig.\ref{fig:relation}, we can re-write the local Q-function into overlapped and non-overlapped effective regions as
\begin{equation}
\begin{aligned}
    Q_{\theta_i}&(s_i, a_i, \{a_{i'}\}_{i'\in\mathcal{B}^{-i}}) =   Q_{\theta_i}(\overline{s}_i, a_i) + \sum_{i'\in\mathcal{B}^{-i}} Q_{\theta_i}(s_{i,i'}, a_i, a_{i'})\\
    &\overset{(a)}{=} \mathds{E}_{a_{i}'\sim \pi_{\theta^{i}}(a_{i}'|s_{i}')}\Big[\sum_{t'=t}^\infty[ r(\overline{s}_i',a_i') + \sum_{i'\in\mathcal{B}^{-i}}\mathds{E}_{a_{i'}'\sim \pi_{\theta^{i'}}(a_{i'}'|s_{i'}')} r(s_{i,i'}', a_i', a_{i'}')] \Big],
\end{aligned}
\label{Q_function_sep}
\end{equation}
where through $(a)$, we expand the Q-function based on the definition of Q-function.
% Similarly, the local state-value function can also be written as
% \begin{equation}
% \begin{aligned}
%     V_{\theta_i}(s_i)& = \mathds{E}_{s_i\sim \mathds{P}_\theta(s_i)}[\mathds{E}_{a_{i}\sim \pi_{\theta_{i}}(a_{i}|s_{i})}Q_{\theta_i}(\overline{s}_i,a_i) \\
%     & + \sum_{i'\in \mathcal{B}^{-i}}\mathds{E}_{a_{i'}\sim \pi_{\theta_{i'}}(a_{i'}|s_{i'})}[Q_{\theta_i}(s_{i,i'}, a_i, a_{i'})]|S_t=s].
% \end{aligned}
% \end{equation}
In this case, both Q-function and value function are composed into stationary term with local information ($\overline{s}_i$ and $a_i$) and cooperation term with neighbors' information ($s_{i,i'}$ and both $a_i$ and $a_{i'}$). The stationary term is dominated by the current agent with negligible influences from other agents. Thus, the precise estimation can be precisely obtained for $Q_{\theta^i}(\overline{s}_i, a_i)$ or $V_{\theta^i}(\overline{s}_i)$. The cooperate term, which depends on neighbor agents' actions or policies, is non-stationary. Such non-stationary environment caused by unknown neighbour's policy is the major reason limiting the performance of our considered multi-agent system. It is hard to model the opponents' policy without any knowledge from opponents \cite{Panait2005Cooperative}. Also, the change of neighbors' actions can easily make previous experience expire, which even degrades the performance. The non-stationary estimation error $\epsilon(\omega^i)$ without action information from neighbouring agents can be written as 
\begin{equation}
\begin{aligned}
    \epsilon(\omega^i) = Q_{\omega^i}(s_{i,i'}, a_i) - \mathds{E}_{a_{i'}\sim\pi_{\theta^{i'}}(a_{i'}|s_{i'})}Q_{\theta^i}(s_{i,i'}, a_i, a_{i'}) \\
    % V_{\delta^i}(s_{i,i'}) &\rightarrow \mathds{E}_{a_{i}\sim\pi_{\theta^{i}}(a_{i}|s_{i})}\mathds{E}_{a_{i'}\sim\pi_{\theta^{i'}}(a_{i'}|s_{i'})}Q_{\theta^i}(s_{i,i'}, a_i, a_{i'}),
\end{aligned}
\label{inconsist}
\end{equation}
where $\omega^i$ is the parameter of the estimator in $i$-th agent for Q-function. 

\subsection{Federated Learning for Non-stationary}
To reduce the estimation error in the non-stationary part, sharing the information to neighbourhoods is the possible way, which can be performed via sharing the features from the agents' network to their neighbours. However, this method requires frequent communication among agents and is time and resource consuming. In this section, we introduce the motivation and benefit of applying federated learning among all agents, where the features are naturally shared with aligned
global models among agents. 

The theoretical basis of applying federated learning in multi-agent RL is from the centralized-training-decentralized-execution scheme, which significantly improves the performance and accelerate the learning process \cite{Peng2021Multi, Nasir2019Multi}. As the environment around agents is similar especially with certain assumptions, i.e. users' locations follows PPP distribution and requests are sent randomly, the knowledge of certain agent can be shared among the network. We decompose this process via a federated learning scheme to distributively train the agents and reduce the backbone traffic. In this way, the distributed model can be trained using local data captured by each agent, which is then aggregated as a global model. We then introduce the potential benefits of federated learning in multi-agent RL.

The estimation error caused by unknown neighbours' action can be reduced by federated learning. During federated process, a combined global model is aggregated from and shared with all agents. With aligned policy $\pi_\theta$ known for all agents, the non-stationary term of estimation error in Eq. \eqref{inconsist} (due to the unknown neighbours' policy) is reduced to
\begin{equation}
\begin{aligned}
    \epsilon(\omega^i) = Q_{\omega}(s_{i,i'}, a_i) - \mathds{E}_{a_{i'}\sim\pi_{\theta}(a_{i'}|s_{i'})}Q_{\theta}(s_{i,i'}, a_i, a_{i'}) \\
\end{aligned}
\label{estimation_error_shared}
\end{equation}
where $\theta=1/|\mathcal{B}|\sum_{i\in\mathcal{B}}\theta^i$ and $\omega=1/|\mathcal{B}|\sum_{i\in\mathcal{B}}\omega^i$ are the global parameters for policy, Q-function, and value function following the federated averaging (FedAvg) algorithm, which simply average the gradient update from agents. With shared parameters in value function, neighbouring agents can obtain the same features from the common observable overlapped effective region, i.e. $s_{i,i'}$, without communication. As illustrated in previous section, the Q-function or value function of each location is only correlates to its local features. Hence, the same features for $s_{i,i'}$ can be obtained by $i$th and $i'$th agents. In this way, a kind of consistency can be maintained among agents by sharing the features of overlaped effective area among neighbours. But it is still not possible to eliminate the estimation error and obtain a precise estimation of Q-function or value function due to the $i$th agent cannot access the the neighboring agent's state $s^{i'}$. 

The remaining estimation error in Eq.\eqref{estimation_error_shared} without the opponent's action input still influences by the unknown state in neighbor's observation:
\begin{equation}
\begin{aligned}
    \overline{\epsilon}(\omega^i) =& \mathds{E}_{a_{i'}\sim\pi_{\theta}(a_{i'}|s_{i,i'})}Q_{\theta}(s_{i,i'}, a_i, a_{i'}) - \mathds{E}_{a_{i'}\sim\pi_{\theta}(a_{i'}|s_{i'})}Q_{\theta}(s_{i,i'}, a_i, a_{i'}) \\
    % \overline{\epsilon}(\delta^i) =& \mathds{E}_{a_{i}\sim\pi_{\theta}(a_{i}|s_{i})}\mathds{E}_{a_{i'}\sim\pi_{\theta}(a_{i'}|s_{i,i'})}Q_{\theta}(s_{i,i'}, a_i, a_{i'})\\
    % &- \mathds{E}_{a_{i}\sim\pi_{\theta}(a_{i}|s_{i})}\mathds{E}_{a_{i'}\sim\pi_{\theta}(a_{i'}|s_{i'})}Q_{\theta}(s_{i,i'}, a_i, a_{i'}).
\end{aligned}
\end{equation}
Such estimation error decreases with increasing portion of overlapped coverage region $s_{i, i'}$ (allowing more features to be shared), which enhances the performance of cooperation. Meanwhile, with small portion of overlapped coverage region $s_{i, i'}$, the error is small and tolerable.
Also, the variance of target function in overlapped coverage region decreases with the increasing number of overlapping agents. This reduces the effectiveness of our considered algorithm with a high degree of connected APs and is the remaining problem of our proposed architecture. The problem can be partly solved by applying distributional estimation with the cost of complexity \cite{Bellemare2017Distributional}.

% The parameter for policy in actor part is updated with gradient and the estimation result from itself:
% \begin{equation}
% \begin{aligned}
%     \nabla_\theta J(\theta) &=  \mathds{E}_{(s_i,a^i)\sim \mathds{P}_\theta}[\nabla_{\theta}\log\pi_\theta(a^i|s_i)(Q_{\omega}(s_i, a^i) - \sum_{a^i\in\mathcal{A}^i}\pi_{\theta^i}(s_i,a^i)Q_\omega(s_i,a^i))],
% \end{aligned}
% \end{equation}
% where the parameter of policy $\theta^i$ in agent $i$ is updated by the sampled state-action pair during training:
% \begin{equation}
%     \overline{\theta}_{t+1}^i \leftarrow \theta_t^i + \alpha^\theta_t \nabla_{\theta^i}\log{\pi_{\theta^i_t}}(s_i_t,a^i_t) (Q_{\omega^i_t}(s_i_t,a^i_t) - \sum_{a^i\in\mathcal{A}^i}\pi_{\theta^i_t}(s_i_t,a^i_t)Q_{\omega^i_t}(s_i_t,a^i_t)).
% \end{equation}

\section{Algorithm}
We present our proposed federated RL algorithm for the partially observable networked multi-agent system in the multi-cell network. This framework can be used to implement various RL approaches. Here, we show the architecture with a policy-based actor-critic RL algorithm, where the Q-function is estimated in the critic part to guide the update of the policy generated by the actor part.

\begin{algorithm}[t]
\setstretch{0.8}
\DontPrintSemicolon
\SetKwData{Left}{left}\SetKwData{This}{this}\SetKwData{Up}{up}\SetKwFunction{Vpsnr}{V-PSNR}\SetKwFunction{Gamend}{Game end}\SetKwFunction{Reset}{Reset}\SetKwFunction{Rot}{Rotation}\SetKwFunction{FedAvg}{FedAvg}\SetKwFunction{PO}{Partially Observe}\SetKwFunction{Eval}{Evaluation}\SetKwInOut{Input}{input}\SetKwInOut{Output}{output}\SetKwFunction{FederatedStep}{Federated Step}

\BlankLine
Initiate environment $Env$, state $s_0$, and the initial values of the parameters $\{\theta^i\}_{i\in\mathcal{B}}$ and $\{\omega^i\}_{i\in\mathcal{B}}$.\; 
\Repeat{Performance Not Improved}{
\If{\Gamend}{
Reset $Env$ and $t=0$, obtain new $S_0$\;
}
\For{$i\in\mathcal{B}$}{
Obtain $O^i_t$ from $S_t$\;
Select an action $A^i_t\sim\pi_{\theta^i_t}(O^i_t)$\;
}
Forms joint action $a_t=(A^i_t)_{i\in\mathcal{B}}$, the environment move to $S_{t+1}$\;
\For{$i\in\mathcal{B}$}{
Observe local reward $r^i_t$ from $S_{t+1}$\;
Update actor's and critic's parameters following Eq.\eqref{r_update} and Eq.\eqref{policy_update} or categorical algorithm and the error from CORAL\;
}
Update global model by averaging $\theta^i_t$ and $\omega^i_t$\;
Update average reward following \eqref{r_update_non_epi}\;
}
  \caption{Federated Reinforcement Learning Algorithm for PONMDP.}
  \label{algo_frl}
\end{algorithm}

For episodic task, we define the parameters for actor and critic in agent $i$ at time $t$ as $\theta^i_t$ and $\omega_t^i$. With joint state $S_t$ and action $A_t$, the update procedure in $i$-th agent for critic network with the TD-error at time instant $t$ follows
\begin{equation}
\begin{aligned}
    \overline{\omega}^i_{t+1} = \omega^i_t &+ \alpha^\omega_t \nabla_\omega Q_{\omega^i_t}(S^i_t,A^i_t,A^{-i}_t) (r^i_{t+1} + Q_{\omega^i_t}(S^i_{t+1},A^i_{t+1},A^{-i}_t) - Q_{\omega^i_t}(S^i_t,A^i_t,A^{-i}_t)),
\end{aligned}
    \label{r_update}
\end{equation}
where $\alpha^\omega_t$ is the step size for critic network. With distributional RL, the TD error is the cross-entropy loss of the KL divergence between the current return and estimated distribution of the return following categorical algorithm in \cite[Algorithm.1]{Bellemare2017Distributional}.
The actor is updated follows
\begin{equation}
    \overline{\theta}^i_{t+1} = \theta^i_t + \alpha^\theta_t \nabla_\theta \log {\pi_{\theta^i}}(O^i_t|S^i_t) Q_{\theta^i_t}(S^i_{t+1},A^i_{t+1},A^{-i}_t),
    \label{policy_update}
\end{equation}
where $\alpha^\theta_t$ is the step size for critic network. 

For non-episodic task ($T<\infty$), the average-reward as $r(\pi_\theta) = \mathds{E}_{(s,a)\sim\mathds{P}_\theta(s,a)}[r(s,a)] = \sum_s d_\theta(s)\sum_a\pi_\theta(a|s)r(s,a)$, and the critic network fit the differential return between rewards and the average reward. The network update with TD-error and estimated average reward $\hat{r}^i_t$ at $i$-th agent follows
\begin{equation}
\begin{aligned}
    \overline{\omega}^i_{t+1} \leftarrow \omega^i_t & + \alpha^\omega_t \nabla_\omega Q_{\omega^i_t}(S^i_t,A^i_t,A^{-i}_t) (r^i_{t+1} - \hat{r}^i_t + Q_{\omega^i_t}(S^i_{t+1},A^i_{t+1},A^{-i}_t) - Q_{\omega^i_t}(S^i_t,A^i_t,A^{-i}_t)).
\end{aligned}
    \label{Q_update_non_epi}
\end{equation}
The average reward is updated via
\begin{equation}
\begin{aligned}
    \hat{r}^i_{t+1} \leftarrow \hat{r}^i_t &+ \alpha^r (r^i_{t+1} - \hat{r}^i_t + Q_{\omega^i_t}(S^i_{t+1},A^i_{t+1},A^{-i}_t) - Q_{\omega^i_t}(S^i_t,A^i_t,A^{-i}_t)),
\end{aligned}
    \label{r_update_non_epi}
\end{equation}
where $\alpha^r$ is reward update parameter. Then, in each federated step, the parameter of critic step is aggregated and averaged
\begin{equation}
\begin{aligned}
    &\omega_{t+1} = 1/|\mathcal{B}|\sum_{i\in\mathcal{B}}\overline{\omega}^i_t 
    \quad\quad \theta_{t+1} = 1/|\mathcal{B}|\sum_{i\in\mathcal{B}}\overline{\theta}^i_t \quad\quad  \hat{r}_{t+1} = 1/|\mathcal{B}|\sum_{i\in\mathcal{B}}\hat{r}^i_{t+1}.
\end{aligned}
\end{equation}

% It should be noted that Q-function can be estimated via $Q_\theta(s_t,a_t) \sim r(s_t,a_t) + V_\theta(s_t,a_t)$. Thus, we only need one network to fit the value function.

% Then, we present our architecture in Algorithm.\ref{algo_frl}. It should be noted that the algorithm is also compatible for updating with value function or an advantage depending on the actual algorithm used to update and fit the network, e.g. actor-critic. The advantage value is an unbiased relative indicator that gives the value of whether the current action is better or worse than the average. 

% This is achieved by adding a baseline which can be an average value of reward $\overline{r}$ or state-value function. The state-value function is denoted by the expectation of actions over the state-action function, i.e. $V_{\theta}(S_t) = \mathds{E}_{A_t\sim\pi_\theta(A_t|S_t)}Q_\theta(S_t,A_t)$. And the resulting indicator is called advantage function, which is denoted as $A_{\theta}(S_t, A_t) = Q_\theta(S_t,A_t) - V_\theta(S_t)$.

\subsection{Converge Condition}
The convergence can be proof via Kushner-Clark Lemma, which gives four conditions for the convergence of ordinary differential equations. We first make several basic assumptions, which aligns with the neural network properties.
\begin{assumption}
For the agent $i$, the Q-function can be written as a combination of features from independent locations: $Q^i(s_i,a^i,a^{-i}) = \omega^\top \phi(s_i,a^i,a^{-i})$, where $\phi(s_i,a^i,a^{-i}) = [\phi^1(s_i,a^i,a^{-i}), ...$ $,\phi^K(s_i,a^i,a^{-i})]^\top\in \mathds{R}^K$. The feature matrix $\Phi=[\phi(s_i,a^i,a^{-i})]\in\mathds{R}^{|\mathcal{S}||\mathcal{A}^i||\mathcal{A}^{-i}|\times K}$ is uniformly bounded and full rank, which means 0 is not a eigenvalue of $\Phi$ (there does not exist a vector $v\in \mathds{R}^K$ which gives $\phi v=\mathds{1}$).
\label{assumption_feature}
\end{assumption}
As illustrated in previous sections, the features in the Q/value function are independent, which only depends on local features.
% In this way, we have $\nabla_{\omega} Q^i(s_i,a^i,a^{-i}) = \phi$.

Following the convergence proof of single agent actor-critic algorithm, the update rate of $\omega$ and $\theta$ should follow the condition in
\begin{assumption}
The update rate of $\alpha^\omega$ and $\alpha^\theta$ satisfy
$$\sum_{t}\alpha^\omega_t = \sum_{t}\alpha^\theta_t = \infty,\ \sum_{t}(\alpha^\omega_t)^2 + \sum_{t}(\alpha^\theta_t)^2 \leq \infty$$
\label{update_rate_assumption_rate}
\end{assumption}

We also introduce the assumption for the reward, value and Q function, which are commonly used in RL.
\begin{assumption}
The value and Q function is separable, and the local Q and value functions are L-Lipschitz continues while each local function has bounded support.
\label{update_rate_assumption_lipschitz}
\end{assumption}

Then, to show the convergence of our proposed federated algorithm, we first analyze the critic step's convergence while assuming a fixed policy $\pi_\theta$ following the two-time-scale SA analysis \cite{Konda2004Convergence}. The convergence of actor step upon converged critic is nicely shown by literature.
% To simplify the notations, we write the TD-error in agent $i$ at time $t$ as $\delta^i_t = r^i_{t+1} + Q(s_i_{t+1},a_i_{t+1}, a^{-i}_{t+1}) - Q(s_i_{t}, a_i_{t}, a^{-i}_{t})$. Thus, the update of critic can be written as
% \begin{equation}
%     \omega_{t+1} = \omega_t + \frac{1}{|\mathcal{B}|}\alpha^\omega_t\sum_{i\in\mathcal{B}}\delta^i_t\phi^i_t.
% \end{equation}

First, it is possible to consider the critic update via an ordinary differential equation
\begin{equation}
    \dot{z} = \Phi^\top D^{s,a}_\theta r^i(s,a) - \Phi^\top D^{s,a}_\theta(P^\theta-I)\Phi \omega,
    \label{ode_critic}
\end{equation}
where $D^{s,a}_\theta$ is the probability of the existence of state-action pair $(s,a)$, i.e. $D^{s,a}_\theta = \text{diag}[d_\theta(s)\pi_\theta(a|s), s\in\mathcal{S}, a\in\mathcal{A}]$, $P^\theta$ is the transition probability from $(s,a)$ to $(s',a')$ under policy $\theta$, i.e. $P^\theta(s',a'|s,a) = P(s'|s,a)\pi_\theta(s',a')$.
% Thus, with some $K <\infty$, we can have
% \begin{equation}
%     \mathds{E}(||\alpha^\omega_t\frac{1}{|\mathcal{B}|}\sum_{i\in\mathcal{B}}\delta^i_t\phi^i_t ||^2|s_t,a_t,...) \leq ||\mathds{E}(\alpha^\omega_t\frac{1}{|\mathcal{B}|}\sum_{i\in\mathcal{B}}\delta^i_t\phi^i_t |s_t,a_t,...)||^2 \leq K(1+||\omega_t||^2).
% \end{equation}
% Since $1/|\mathcal{B}| \in (0,1]$. The equation can be further bounded over $\sup_t ||\omega_t|| < M$. In this way, we can get:
% \begin{equation}
%     \lim_{t\rightarrow\infty}(\sup_{t}||\alpha^\omega_t\frac{1}{|\mathcal{B}|}\sum_{i\in\mathcal{B}}\delta^i_t\phi^i_t ||^2)<\infty
% \end{equation}
% Recall that we can have $\sup_t||\omega_t||<\infty$ follow the previous assumptions. 

We now justify why the update steps satisfy the Assumptions.1-4 for Kushner-Clark Lemma \cite{Prasad2014Actor}: 1) Since the $\frac{1}{|\mathcal{B}|}\sum_{i\in\mathcal{B}}\delta^i_t$ is the function of $\omega^i_t$, i.e. $\delta^i_t = r^i_{t+1} + (\phi^i_{t+1})^\top\omega_t - (\phi^i_{t})^\top\omega_t$. Then, with uniformly bounded $\phi$, $\delta$ is Lipschitz continuous in $\omega_t$, as all components are linear; 2) $P^\theta$ is a non-negative matrix. According to Perron-Frobenius theorem, $P^\theta$ has one eigenvalue equal to the spectrum-radius of $P^\theta$, whose maximum value is 1 for considered probability transfer matrix. Other eigenvalues are less than 1. Thus, it is possible to have an zero eigenvalue in vector $P^\theta-I$, which gives a vector $v$ that satisfies $\Phi v= \mathds{1}$. However, this special case rarely exists. All eigenvalues are negative real number in $P^\theta-I$. Hence, Eq.\eqref{ode_critic} has a asymptotically stable solution (equilibrium) \cite[Theorem. 2]{Prasad2014Actor} when
\begin{equation}
    \Phi^\top D^{s,a}_\theta [r^i(s,a) - (P^\theta-I)\Phi \omega]=0,
\end{equation}
where the solution $\omega_\theta$ is unique;
3) The step size $\alpha^\omega_t$ has the property in Assumption.\ref{update_rate_assumption_rate}; 4) The federated average operation removes the noisy part which denotes the difference between the local critic models by keeping parameters aligned. Thus, this condition is absent. In this way, the update of the critic part follows the Kushner-Clark Lemma, which converges almost surely when $t\rightarrow\infty$. Thus, we complete the proof of the critic convergence \cite{Shalabh2009Natural}. Then, following the proof of original actor-critic algorithm and two-time-scale SA analysis \cite{Diddigi2019Actor}, the actor part can converge guided by a converged critic, which concludes the proof.
% \cite{Zhang2018Fully}
% \cite{KondaVijay2000Actor,Diddigi2019Actor,Borkar2005actor}

% For actor part, the update is presented as
% \begin{equation}
%     \theta_{t+1} = \frac{1}{|\mathcal{B}|}\sum_{i\in\mathcal{B}}\overline{\theta}_{t+1}^i = \frac{1}{|\mathcal{B}|}\sum_{i\in\mathcal{B}}[\theta_t^i + \alpha^\theta_t \nabla_{\theta^i}\log {\pi_{\theta^i_t}} A^i_t],
% \end{equation}
% where the advantage $A^i_t$ is defined as $A^i_t = Q(s_i_t,a^i_t,a^{-i}_t|\omega^i_t) - \sum_{a^i\in\mathcal{A}^i}\pi_{\theta^i_t}(s_i_t,a^i_t,a^{-i}_t)Q(s_i_t,a^i_t,a^{-i}_t|\omega^i_t)$.

% The expectation of the actor update at time $t$ can be written as
% \begin{equation}
%     \Delta \theta = \frac{1}{|\mathcal{B}|}\sum_{i\in\mathcal{B}} \sum_{s_t\in\mathcal{S},a_t\in\mathcal{A}}d_{\theta}(s_t)\pi_\theta(s_t,a_t)\nabla_{\theta^i}\log {\pi_{\theta^i_t}} A^i_t,
% \end{equation}
% where $\Delta \theta$ is a continues function in $\theta^i_t$. Since $\nabla_{\theta^i}\log {\pi_{\theta^i_t}}$ is continuous as Assumption. $d_{\theta}(s_t)\pi_\theta(s_t,a_t)$ is continues in $\theta^i_t$ since the distribution is stationary. $A^i_t$ is continues in $\theta^i_t$ since $\omega_\theta$ is the unique solution to the linear equation \ref{ode_critic}. 

\subsection{Convergence Speed Analysis with Informational Model}
Normally in federated learning, the federated operation is required to be performed every learning step (FedAvg). However, it is resource and time consuming to transmit the entire model each time, which is also impossible for a specific communication system. But reducing the federated frequency can significantly lower the accuracy and convergence speed in conventional classification tasks. Thus, it is necessary to analyse the effect of federated frequency on the convergence speed of our proposed architecture. In this section, we applied the informational model for multi-agent learning defined in \cite{Terry2020Revisiting} and extend it to federated cases to derive the upper bound for converging speed under different federated frequencies for our model.

Similar to Eq.\eqref{Q_function_sep}, we can separate the knowledge or information required to fit the Q-function in each agent $i$ into local information (information in $\overline{s}^i$) and cooperative information (information in $s^{i,i'} \forall i'\in I^{-i}$). 
We also define the local information in $i$-th agent at time $t$ as $\mathcal{I}_{i,\text{env}}(t)$ and the cooperative information between $i$ and its neighbor $i'$ as $\mathcal{I}_{i,i'}(t)$. In this way, we have the overall information in $i$th agent at time $t$ as
\begin{equation}
    \mathcal{I}_{i}(t) = \mathcal{I}_{i,\text{env}}(t) + \sum_{i'\in \mathcal{B}^{-i}}\mathcal{I}_{i,i'}(t).
\end{equation}

During the learning procedure, the information increases over each time step. For any agent $i$ in a group of agent $\mathcal{B}$ with neighbor agents $\mathcal{B}^{-i}$, the information gain in each learning time step is defined as
\begin{equation}
    \Delta^{\uparrow}\mathcal{I}_i(t) = \Delta^{\uparrow}\mathcal{I}_{i,\text{env}}(t) + \sum_{i'\in \mathcal{B}^{-i}}\Delta^{\uparrow}\mathcal{I}_{i,i'}(t),
    \label{InfoGeneral}
\end{equation}
where $\Delta^{\uparrow}\mathcal{I}_{i,\text{env}}$ is the gain for local information, $\Delta^{\uparrow}\mathcal{I}_{i,i'}$ is the gain for cooperation information between agent $i$ and its neighbor $i'$. For any agent $i$ in a group of agent $I$ with neighbor agents $I^{-i}$, the local information required to converge is defined as $\mathcal{C}_{i,\text{env}}$ and the cooperative information between $i$ and neighbor $i'$ is $\mathcal{C}_{i,i'} \forall i'\in I^{-i}$. These two terms satisfy
\begin{equation}
    \mathcal{C}_{i,\text{env}} + \sum_{i'\in I^{-i}}\mathcal{C}_{i,i'} = 1, \mathcal{C}_{i,\text{env}}\in[0,1], \mathcal{C}_{i,i'}\in[0,1].
\end{equation}
To model the value of information gain, we denote the function of the information gain learnt as $\Lambda$, which is a function of the rest of information. Then, the information gain for local information and cooperation information can be written as
\begin{equation}
    \begin{aligned}
    \Delta^{\uparrow}\mathcal{I}_{i,\text{env}}(t) & = \mathcal{K}_{i, \text{env}} \Lambda\left(\mathcal{C}_{i, \text{env}}-\mathcal{I}_{i, \text{env}}(t-1)\right),\text{ and}
    \end{aligned}
    \label{InfoDetail_env}
\end{equation}
\begin{equation}
    \begin{aligned}
    \Delta^{\uparrow}\mathcal{I}_{i,i'}(t) & = \mathcal{K}_{i, i'} \Lambda\left(\mathcal{C}_{i, i'}-\mathcal{I}_{i, i'}(t-1)\right), \\
    \end{aligned}
    \label{InfoDetail_coop}
\end{equation}
respectively, where the value of $\mathcal{K}_{i,\text{env}}\in[0,1]$ and $\mathcal{K}_{i,i'}\in[0,1]$ refer to the learning rate coefficient, which corresponds to the several settings in the algorithm, such as batch size, learning rate, and etc, and may differ among agents. It should be noted that the learning function has the property as $\Lambda(x)\leq x$, since the learnt information can't be larger than the rest unlearnt information.

As illustrated, the change of neighbors' policy can make the previous learnt information outdated. The information loss is highly correlated to the amount of new information learnt by neighbor agents, which is unknown for current agent. As the local information $\overline{s}^i$ can be seen as stationary so there is no information loss in the learning the local information part. For the cooperation information, we define the information loss between agent $i$ and $i'$ as
\begin{equation}
    \Delta^{\downarrow} \mathcal{I}_{i,i'}(t)=\frac{\Delta^{\uparrow} \mathcal{I}_{i'}(t)}{\mathcal{I}_{i'}(t-1)+\Delta^{\uparrow} \mathcal{I}_{i'}(t)} \mathcal{I}_{i, i'}(t-1).
    \label{DeltaDown}
\end{equation}

Combining Eq.\eqref{InfoGeneral} and Eq.\eqref{DeltaDown}, we denote the information gain $\Delta \mathcal{I}_{i}(t)$ for agent $i$ from time $(t-1)$ to $t$ as
\begin{equation}
    \Delta \mathcal{I}_{i}(t) = \Delta^{\uparrow}\mathcal{I}_{i,\text{env}}(t) + \sum_{i'\in \mathcal{B}^{-i}}(\Delta^{\uparrow}\mathcal{I}_{i,i'}(t) - \Delta^{\downarrow} \mathcal{I}_{i,i'}(t)).
\end{equation}

With the help of federated learning, agents can share the information learnt locally among the group of agents. Moreover, after the agents share the same learning model after federated step, it has full information for the neighbor agents in the next learning step. In this way, there is no information loss after each federated average operation. Thus, the information gain for local update step in $t$th agent is denoted as
\begin{equation}
    \Delta \mathcal{I}_i(t) = \Delta^{\uparrow}\mathcal{I}_{i,\text{env}}(t) + \sum_{i'\in \mathcal{B}^{-i}}(\Delta^{\uparrow}\mathcal{I}_{i,i'}(t) - \mathds{1}[t|F]\Delta^{\downarrow}\mathcal{I}_{i,i'}(t)),
    \label{FLEachInfo}
\end{equation}
where $\mathds{1}[t|F]=0$ when $t$ can be fully divided by $F$ and the federated average is performed every $F$ local learning steps.

When performing federated learning, the information learnt by all agents are shared and added up among agents. Thus, the information gain in federated step after $F-1$ local update in $i$th agent is denoted as
\begin{equation}
\begin{aligned}
    \Delta &\mathcal{I}_i(t) = \Delta^{\uparrow}\mathcal{I}_{i,\text{env}}(t) + \sum_{i'\in \mathcal{B}^{-i}}(\Delta^{\uparrow}\mathcal{I}_{i,i'}(t) - \mathds{1}[t|F]\Delta^{\downarrow}\mathcal{I}_{i,i'}(t)) \\
    & + \sum_{t'=t-F+1}^{t}\sum_{i'\in \mathcal{B}/{i}}\Delta^{\uparrow}\mathcal{I}_{i',\text{env}}(t') + \sum_{t'=t-F+1}^{t}\sum_{i'\in\mathcal{B}/i}\sum_{j'\in\mathcal{B}^{-i'}}(\Delta^{\uparrow}\mathcal{I}_{i',j'}(t') - \mathds{1}[t'|F]\Delta^{\downarrow}\mathcal{I}_{i',j'}(t')).
\end{aligned}
    \label{FLEachAdd}
\end{equation}

\begin{assumption}
To simplify the model, we assume agents are co-located in the same pattern with identical environment. The initial amount of information in agents are the same. For simplicity, we denote the information loss between any pair of agents $i$ and $i'$ as $\Delta^{\downarrow}\mathcal{I}_{*,*}(t)$, i.e. $\mathcal{I}_{i,\text{env}}(t)=\mathcal{I}_{j,\text{env}}(t)=\mathcal{I}_{*,\text{env}}(t)$ and  $\mathcal{I}_{i,i'}(t)=\mathcal{I}_{j,j'}(t)=\mathcal{I}_{*,*}(t),\forall i,i',j,j'\in I, i\neq j$. Besides, the information gain in each agent is also assumed to be homogeneous, which significantly reduce the complexity of our analysis. But it reduce generalization for our analysis \cite{Terry2020Revisiting}.
\label{identical_agents}
\end{assumption}

Following Assumption.\ref{identical_agents} and Eq.\eqref{FLEachAdd}, the overall gain in $F-1$ agent's cooperation information update and the following federated update can be written as
\begin{equation}
\begin{aligned}
    \mathcal{I}_i(t) &- \mathcal{I}_i(t-F) = |\mathcal{B}|\sum_{t'=t-F+1}^t\Delta^{\uparrow}\mathcal{I}_{*,\text{env}}(t')  + |\mathcal{B}||\mathcal{B}^{-i}|\sum_{t'=t-F+1}^t(\Delta^{\uparrow}\mathcal{I}_{*,*}(t') - \mathds{1}[t'|F]\Delta^{\downarrow}\mathcal{I}_{*,*}(t')),
\end{aligned}
    \label{FLInfo}
\end{equation}
where the convergence speed is decided by local information ($\mathcal{I}_{*,\text{env}}$) and cooperation information ($\mathcal{I}_{*,*}$) separately. In the following, we analyse the convergence speed of aforementioned two parts individually, and the convergence speed is decided by the larger one between these two results.

First, we analyse the convergence speed of cooperation information part. By substituting Eq.\eqref{InfoDetail_coop} into Eq.\eqref{FLInfo}, the part of information gain for $i$th agent at time $t$ is denoted as
\begin{equation}
\begin{aligned}
    &\Delta^{\uparrow} \mathcal{I}_i(t) = \Delta^{\uparrow} \mathcal{I}_{*,\text{env}}(t) + |\mathcal{B}^{-*}|\Delta^{\uparrow} \mathcal{I}_{*,*}(t)= \mathcal{K}_{\text{env}}\Lambda(\mathcal{C}_\text{env}-\mathcal{I}_{*,\text{env}}(t-1)) + |\mathcal{B}^{-*}|\mathcal{K}_*\Lambda(\mathcal{C}_*-\mathcal{I}_{*,*}(t-1)).
\end{aligned}
\label{FLInfo_gain}
\end{equation}
Similarly, we can expand the cooperation information loss between agent $i$ and $i'$ as Eq.\eqref{information_loss} \cite{Terry2020Revisiting}.
\begin{equation}
\begin{aligned}
    \Delta^{\downarrow} &\mathcal{I}_{*,*}(t) = \frac{|\mathcal{B}^{-*}|\mathcal{K}_*\Lambda(\mathcal{C}_*-\mathcal{I}_{*,*}(t-1))}{\mathcal{I}_{i'}(t-1)+ |\mathcal{B}^{-*}|\mathcal{K}_*\Lambda(\mathcal{C}_*-\mathcal{I}_{*,*}(t-1))} \mathcal{I}_{*,*}(t-1). \\
\end{aligned}
\label{information_loss}
\end{equation}
Substituting Eq.\eqref{information_loss} and cooperation information term in Eq.\eqref{FLInfo_gain} into Eq.\eqref{FLInfo}, the overall gain for cooperation information $\mathcal{I}_{*,*}$ between federated average operations (including $F-1$ local update and a federated update) can be denoted as Eq.\eqref{information_gain_F}.
\begin{equation}
\begin{aligned}
    &\mathcal{I}_{*,*}(t) - \mathcal{I}_{*,*}(t-F) =|\mathcal{B}|\sum_{t'=t-F+1}^t\big(\Delta^{\uparrow} \mathcal{I}_{*,*}(t') - \mathds{1}[t|F]\Delta^{\downarrow} \mathcal{I}_{*,*}(t')\big) \\
    & \leq |\mathcal{B}|\sum_{t'=t-F+2}^t\Big[(\mathcal{K}_*(\mathcal{C}_*-\mathcal{I}_{*,*}(t'-1)))(1-\frac{\mathcal{I}_{*,*}(0)}{\mathcal{C}_{*,\text{env}}/|\mathcal{B}^{-*}| + \mathcal{K}_*\mathcal{C}_* + \mathcal{C}_*})\Big] +  |\mathcal{B}|\mathcal{K}_*(\mathcal{C}_*-\mathcal{I}_{*,*}(t-F)).
\end{aligned}
\label{information_gain_F}
\end{equation}
Then, we denote $t_F$ as the closest time instance with federated average operation from $t$, i.e. $t-t_F < F$. By continuing decomposing $F$ steps iteratively 
between two federated steps with the Eq.\eqref{information_gain_F} with Eq.\eqref{InfoDetail_coop} and Eq.\eqref{information_loss}, we can get Eq.\eqref{information_gain}. 
\begin{equation}
\begin{aligned}
    \mathcal{I}_{*,*}(t) - \mathcal{I}_{*,*}(t_F) &\leq |\mathcal{B}|\sum_{t'=t_F+2}^t(\alpha\mathcal{C}_* - \alpha \mathcal{I}_{*,*}(t'-1)) + |\mathcal{B}|\mathcal{K}_*(\mathcal{C}_*-\mathcal{I}_{*,*}(t_F)) \\
    & = |\mathcal{B}| \alpha (t-t_F-1)\mathcal{C}_* - |\mathcal{B}|\alpha \Big[ \sum_{t'=t_F+2}^{t-1}\mathcal{I}_{*,*}(t'-1) \\
    &\quad+ \mathcal{I}_{*,*}(t-2) + (\alpha\mathcal{C}_*-\alpha \mathcal{I}_{*,*}(t-2)) \Big] + |\mathcal{B}|\mathcal{K}_*(\mathcal{C}_*-\mathcal{I}_{*,*}(t_F)) \\
    & = |\mathcal{B}|[(1-\mathcal{K}_*)(1-(1-\alpha)^{F-1})+\mathcal{K}_*](\mathcal{C}_* - \mathcal{I}_{*,*}(t_F)).
\end{aligned}
\label{information_gain}
\end{equation}

As the federated round $F$ is small compared to the overall learning rounds, we only look at the time after each federated average operation. Then, by further expanding the equation to $t=0$, we have the formulation of $\mathcal{I}_{*,*}(t)$ as Eq.\eqref{information_all}. 
\begin{equation}
\begin{aligned}
    &\mathcal{I}_{*,*}(t) \leq \mathcal{C}_* - (1-|\mathcal{B}|[(1-\mathcal{K}_*)(1-(1-\alpha)^{F-1})+\mathcal{K}_*])^{\lfloor t/F\rfloor }(\mathcal{C}_*-\mathcal{I}_{*,*}(0)),
\end{aligned}
\label{information_all}
\end{equation}
where $\alpha = \mathcal{K}_*(1-\frac{\mathcal{I}_{*,*}(0)}{\mathcal{C}_{*,\text{env}}/|\mathcal{B}^{-*}| + \mathcal{K}_*\mathcal{C}_* + \mathcal{C}_*})$.

We solve $t$ for the upper bound $t$ when $\mathcal{I}_{*,*}(t)\leq \mathcal{C}_*(1-\epsilon)$. We derive our upper-bound for neighbours' part as Eq.\eqref{neighbor_upb}.
\begin{equation}
\begin{aligned}
    t^* = \frac{F\log \frac{\mathcal{C}_*\epsilon}{\mathcal{C}_* - \mathcal{I}_{*,*}(0)}}{\log \Big[ 1 - |\mathcal{B}|\big[ (1-\mathcal{K}_*)(1-(1-\mathcal{K}_* \quad (1 - \frac{\mathcal{I}_{*,*}(0)}{\mathcal{C}_{*,\text{env}}/|\mathcal{B}^i| + \mathcal{K}_*\mathcal{C}_* + \mathcal{C}_*}))^{F-1}) + \mathcal{K}_*\big]\Big]}
\end{aligned}
\label{neighbor_upb}
\end{equation}

Similarly, for local information $\mathcal{I}_{*,\text{env}}$, we have the upper bound of its convergence speed as
\begin{equation}
\begin{aligned}
    \mathcal{I}_{*,\text{env}}(t) &\leq \mathcal{I}_{*,\text{env}}(t-1) + \Delta \mathcal{I}_{*,\text{env}}(t) = (1-|\mathcal{B}|\mathcal{K}_*)\mathcal{I}_{*,\text{env}}(t-1) + |\mathcal{B}|\mathcal{K}_*\mathcal{C}_\text{env} \\
    & \leq \mathcal{C}_\text{env} - (1-|\mathcal{B}|\mathcal{K}_*)^t(\mathcal{C}_\text{env}-\mathcal{I}_{*,\text{env}}(0)). \\
\end{aligned}
\end{equation}

Thus, for $\mathcal{I}_{*,\text{env}}(t)\leq \mathcal{C}_\text{env}(1-\epsilon)$, we have
\begin{equation}
    t_{\text{env}} = \log_{1-|\mathcal{B}|\mathcal{K}_\text{env}} (\frac{\mathcal{C}_\text{env}\epsilon}{\mathcal{C}_\text{env} - \mathcal{I}_{*,\text{env}}(0)}).
    \label{environment_upb}
\end{equation}

The upper bound of convergence time $t$ for agent is the larger one in Eq.\eqref{environment_upb} and Eq.\eqref{neighbor_upb}, which ensures the amount of the information learnt by the agent larger than the threshold $\epsilon$. Note that the $F$ should be small compared to $(\mathcal{C}_* - \mathcal{I}_{*,*})/\mathcal{K}_*$.

\begin{figure}[!htbp]
    \centering
    \includegraphics[width=0.5\textwidth]{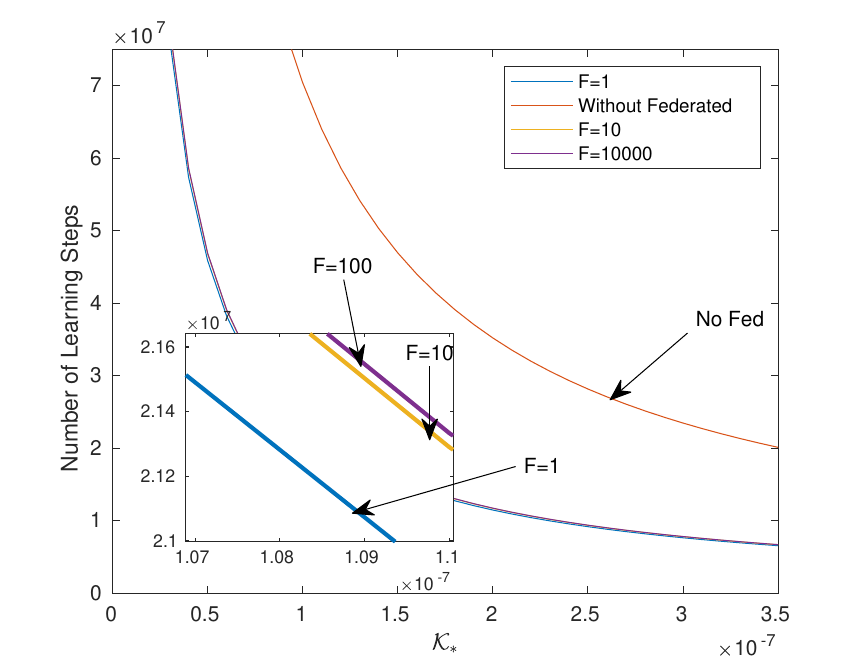}
    \caption{Converge Rate Over $\mathcal{K}_*$ with Different $F$.}
    \label{fig:convergespeed}
\end{figure}

Thus, we have the upper bound of converge time for federated RL with the help of multi-agent information model, which is the maximum number within Eq.\eqref{environment_upb} and Eq.\eqref{neighbor_upb}.
% \begin{equation}
%     t = \max \Big[ \log_{1-|\mathcal{B}|\mathcal{K}_\text{env}} (\frac{\mathcal{C}_\text{env}\epsilon}{\mathcal{C}_\text{env} - \mathcal{I}_{*,\text{env}}(0)}), 
%     F\frac{ \log \frac{\mathcal{C}_*\epsilon}{\mathcal{C}_* - \mathcal{I}_{*,*}(0)}}{\log \Big[ 1 - |\mathcal{B}|\big[ (1-\mathcal{K}_*)(1-\mathcal{K}_* (1 - \frac{\mathcal{I}_{*,*}(0)}{\mathcal{C}_{*,\text{env}}/|\mathcal{B}^i| + \mathcal{K}_*\mathcal{C}_* + \mathcal{C}_*}))^{F-1} + \mathcal{K}_*\big]\Big]} \Big]
%     \label{convergespeed}
% \end{equation}
To visualized the upper-bound and its relations between $F$, We follow the parameter setting with the $\mathcal{C}_{*, \text{env}}=0.1$, $|\mathcal{B}|=10$, $\mathcal{I}_{*,*}(0)=0.01$, and $\epsilon=0.001$ in \cite{Terry2020Revisiting}. We plot the Fig.\ref{fig:convergespeed} which shows that the high federated averaging frequency can significantly reduce the required learning steps, and federated averaging can still significantly accelerate convergence even with relative large $F$.

\subsection{Centralized-Decentralized Mismatch}
We motivate the use of federated learning in the multi-agent RL with the assumption of homogeneous devices, where the environment around agents is assumed to be similar. However, due to the different geometry characteristics in a deployed environment, the \say{averaged} federated model may not always be generalized enough to handle the heterogeneity environment. The heterogeneous data is recognised as one of the major problems in \say{averaged} federated learning when compared with the centralized training approach. The sub-optimality or local characteristic of certain agent's policy can propagate through the federated process and negatively affect other agents' performance, called \say{Centralized-Decentalized Mismatch} \cite{Wang2020Policy}. In our considered CoMP case, some APs locate near rivers or other hidden environment objects limit the visit of users, whose experience is biased. In our simulation environment, the AP at the edge of the network has a very limited choice of cooperation, whose experience is highly personalized and not suitable to be fully accepted by other agents. Moreover, in reality, users are not distributed evenly in the serving range of each agent during the service with certain hot spots (distributed following PCP instead of PPP). Thus, it is important to balance local knowledge and shared knowledge with heterogeneous agents with the centralized-decentralized-mismatch. 

We try to solve this problem via personalizing in our architecture. Personalizing allows each agent to keep its local characteristics while still sharing some global knowledge \cite{Smith2018CoCoA}. We adopt the transfer learning method to realise the personality and reduce the negative effect of centralized-decentralized mismatch in our federated architecture. By bringing the local model similar to the global model when updating the mode, CORelation ALignment (CORAL) is a simple approach for unsupervised model alignment, which minimises the model shift between global model and local model \cite{Sun2016Correlation}. By minimising the second-order statistics between local features and global features, the CORAL helps the local agent to reduce the local loss and adapt part of the global knowledge. Then, the global knowledge can be transferred into the local model while keeping the characteristic of local features. We apply CORAL to our architecture where the loss of CORAL is measured between the output of linear layers before the softmax layers \cite{Sun2016Correlation}.

\section{Simulation Results}
In this section, we provide simulation results to show the effectiveness of our proposed federated multi-agent RL frameworks and verify several architecture designs with our CoMP example case, which requires super high flexibility and scalability. 

We consider a $182m\times 168m$ serving area with $|\mathcal{U}|=160$ PCP distributed users in $10$ clusters with largest radius of \SI{40}{\meter}. The new users' position is generated every time slot and the old users stay on-grid for two time-slots before the service time-out. The users are served by $|\mathcal{B}| = 5\times 4 = 20$ APs. The APs are distributed in cellular with 6 neighbours. The gap between neighbouring APs is $44.3m$ or $52m$. The number of possible cooperation actions is $|\mathcal{A}=12|$, as the maximum size of the cluster is considered as $3$. The APs choose one or two of its neighbour to cooperate. Users' locations follow PCP distribution with $10$ clusters in the area. 

\begin{figure}[t]
    \begin{minipage}[t]{0.48\textwidth}
    \centering
    \includegraphics[width=\textwidth]{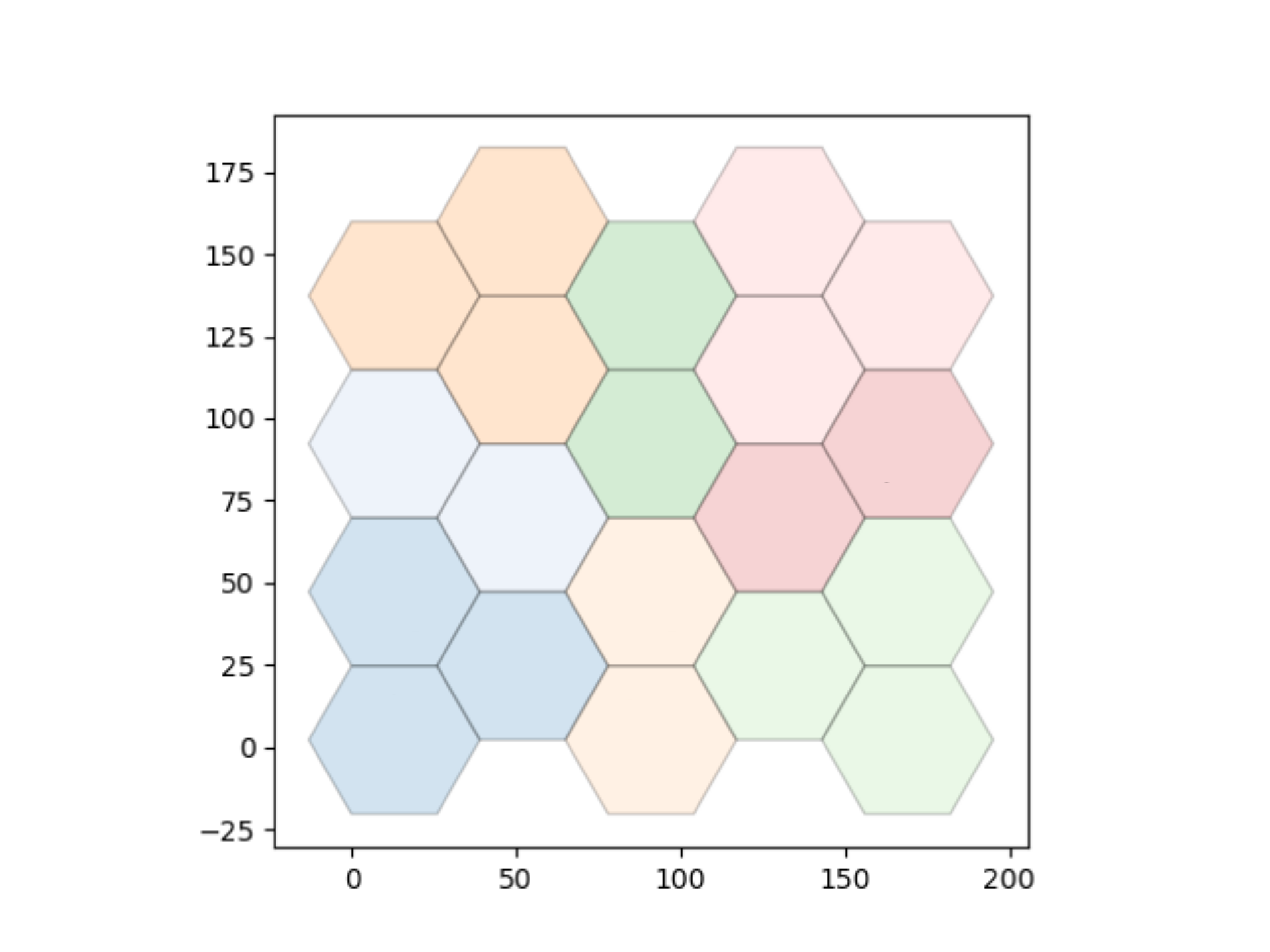}
    \caption{Fixed cooperation scheme for $20$ APs or the learning policy under PPP user distribution.}
    \label{fig:fixcoop}
    \end{minipage}
    \hspace*{0.2cm}
    \begin{minipage}[t]{0.48\textwidth}
    \centering
    \includegraphics[width=\textwidth]{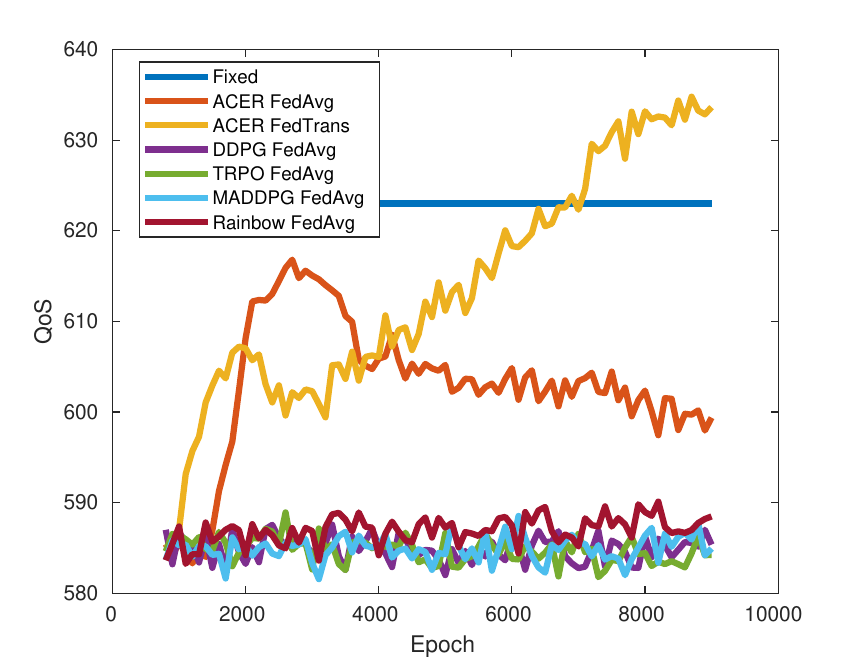}
    \caption{QoS Performance of different RL algorithms under our introduced architecture with FedAvg.}
    \label{fig:methods_performance}
    \end{minipage}
\end{figure}

For the baseline, we use the greedy scheme or fixed cooperation scheme. As we consider the cooperation loss by a maximum cluster size in our model, the cooperation can only bring gain to system performance. Thus, the greedy scheme will lead to a fixed cooperation scheme with maximum cooperating APs, which can be seen as a greedy optimum. The cooperation scheme is presented in Fig.\ref{fig:fixcoop}, where the APs with the same colour are cooperating. We also use the random scheme as a performance baseline, where agents randomly select its actions.

For the interaction between environment and learning algorithm, all agents observe the users' location and requests inside its observation range and make decisions. The algorithms learn from the environment continually without resetting to some default states, i.e. non-episodic. As we illustrated before, the RL algorithms showed significantly different performances in the episodic and non-episodic environments. Empirically, the episodic environment eases the learning difficulty, which is shown in Fig.\ref{fig:methods_performance}. There is limited analysis for the reason for this phenomenon \cite{CoReyes2020Ecological, Naik2019Discounted}. 

For the neural network design, we adopt a three convolutional layer neural network to capture the geometry correlation between APs and users. The network takes a picture, whose value of the pixel presents the existence of neighbour APs or users in the corresponding location. The value is added and normalized if multiple users are located in the same pixel. In this way, the network takes the users' geometry information with constant input size. After the input is processed as a hidden vector by the convolutional layer. The hidden vector then is then used to generate the Q-function estimation by a distributional RL structure with three noisy linear layers \cite{Bellemare2017Distributional}. The noisy linear layers add noise into the result for state-based exploration \cite{Hessel2017Rainbow}. The distributional RL structure allows the estimation to be performed on the value following certain distribution precisely, which is suitable for wireless communication cases. The CORAL layer is added at the end of the network to perform transfer learning.

For the detailed RL algorithms, we show the performance of different RL approaches with our architecture in an example CoMP environment. In Fig.\ref{fig:methods_performance}, we consider actor-critic with experience relay with federated averaging or transfer learning method (ACER FedAvg/FedTrans) with value function in critic, actor-critic with experience relay with federated averaging or transfer learning method (ACER-Q FedAvg) with Q-function in critic, Deep Deterministic Policy Gradient (DDPG) \cite{Lillicrap2015Continuous}, Multi-Agent Deep Deterministic Policy Gradient (MADDPG) \cite{Lowe2017Multia}, Trust Region Policy Optimization (TRPO) \cite{Schulman2015Trust, Gupta2017Cooperative}, and Rainbow \cite{Hessel2017Rainbow} \footnote{The authors acknowledge the use of the research computing facility at King’s College London, Rosalind (\url{https://rosalind.kcl.ac.uk}). The code for this paper is available in \url{https://github.com/paperflight/Fed-MF-MAL/tree/main}.\label{rosalind}}. All methods share a similar size of neural network with similar computation complexity. Our result shows that only ACER FedAvg and ACER FedTrans algorithms show high QoS performance and fast converge speed in our considered CoMP case, while other algorithms failed to converge or converge slowly. Though, the performance of different RL algorithms varies based on the characteristics of different tasks. We apply ACER FedAvg/FedTrans algorithm for later analysis.

\paragraph{Complexity Analysis}
The size of this association problem in our defined environment is $12^{20}$, which is over $4\times 10^{22}$ and impossible to be captured by any existing centralized learning approach. By leveraging the advantage of communication environment, mean-field theory, and graph neural network, our architecture decomposes the problem geometrically and degrade the system complexity from $\mathcal{O}(|\mathcal{A}|^{|\mathcal{B}|})$ to $\sum_{|\mathcal{B}|}\mathcal{O}(|\mathcal{A}|)$ and can be applied in computation resource and memory limited devices. These designs significantly reduce the complexity of cooperative algorithms in large scale networks. The capability of solving this problem can prove the scalability of our introduced architecture.

\begin{figure}[htbp!]
    \begin{minipage}[t]{0.48\textwidth}
    \centering
    \includegraphics[width=\textwidth]{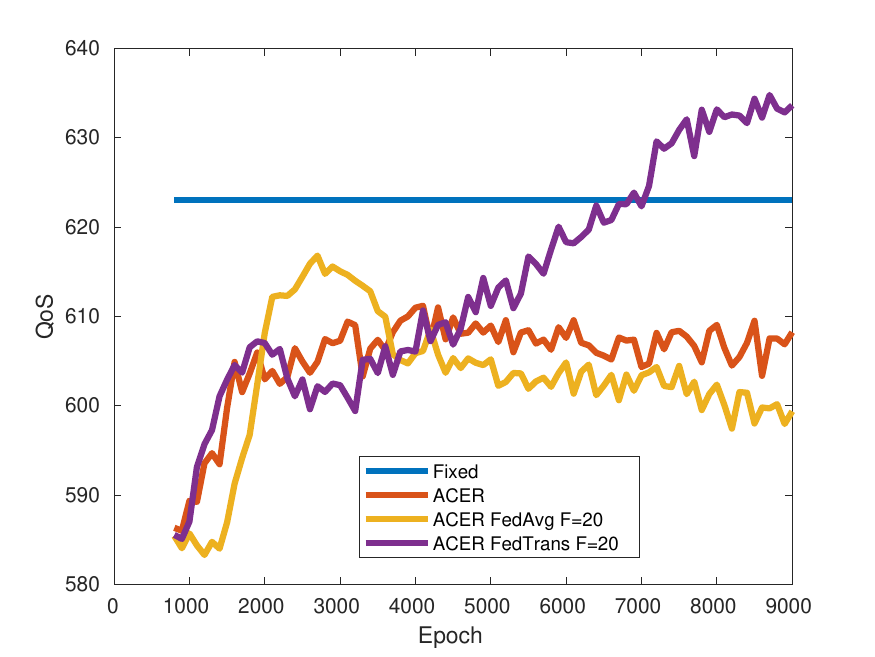}
    \caption{QoS Performance of ACER algorithm with FedAvg algorithm or FedTrans under our introduced architecture.}
    \label{fig:average_steg}
    \end{minipage}
    \hspace*{0.2cm}
    \begin{minipage}[t]{0.48\textwidth}
    \centering
    \includegraphics[width=\textwidth]{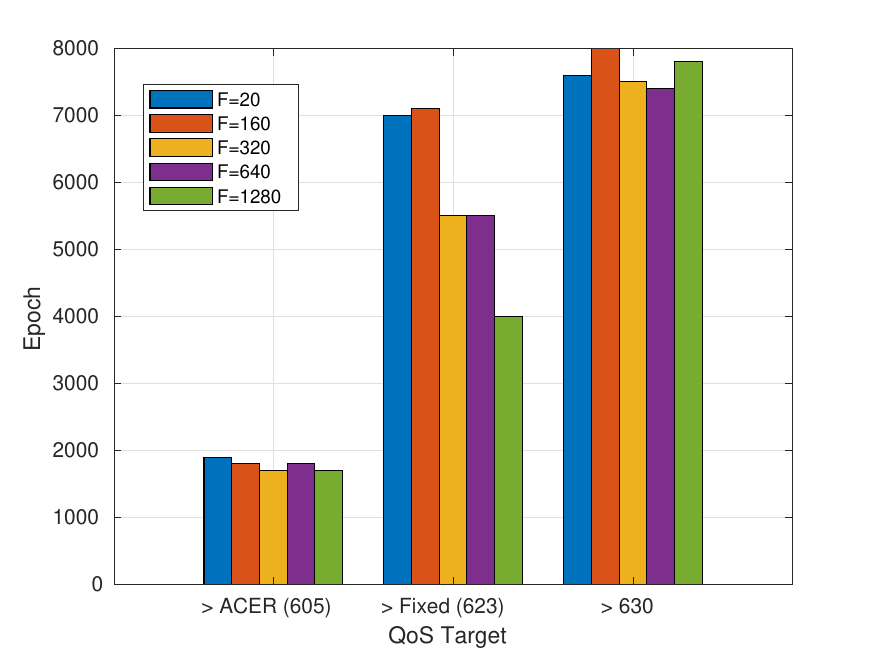}
    \caption{The influence of federated frequency on the performance of ACER algorithm with FedTrans.}
    \label{fig:federated_steps}
    \end{minipage}
\end{figure}

\begin{figure}[t]
    \begin{minipage}[t]{0.48\textwidth}
    \centering
    \includegraphics[width=\textwidth]{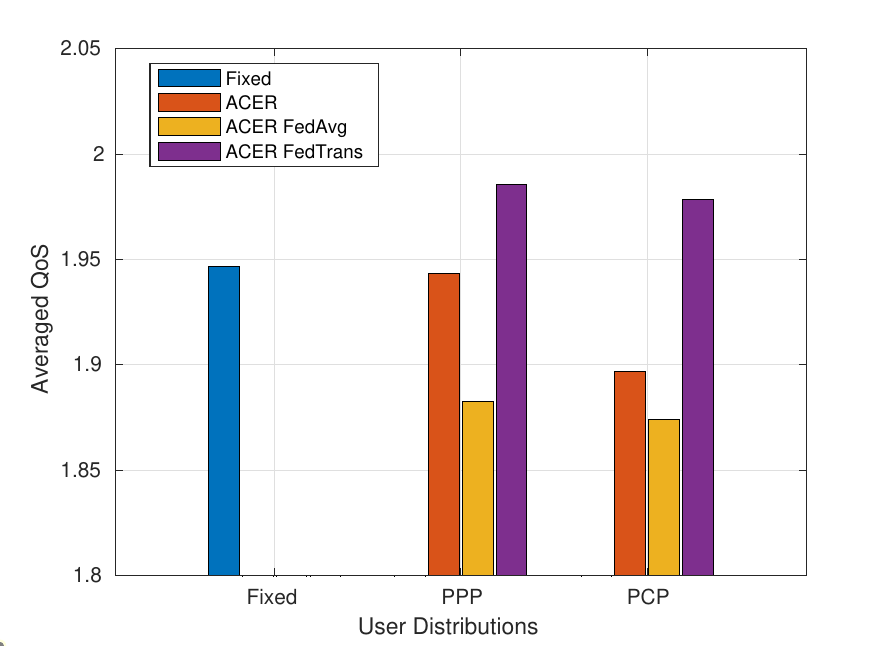}
    \caption{The difference in averaged QoS performance for ACER algorithms with PPP distributed users or PCP distributed users.}
    \label{fig:distribution}
    \end{minipage}
    \hspace*{0.2cm}
    \begin{minipage}[t]{0.48\textwidth}
    \centering
    \includegraphics[width=\textwidth]{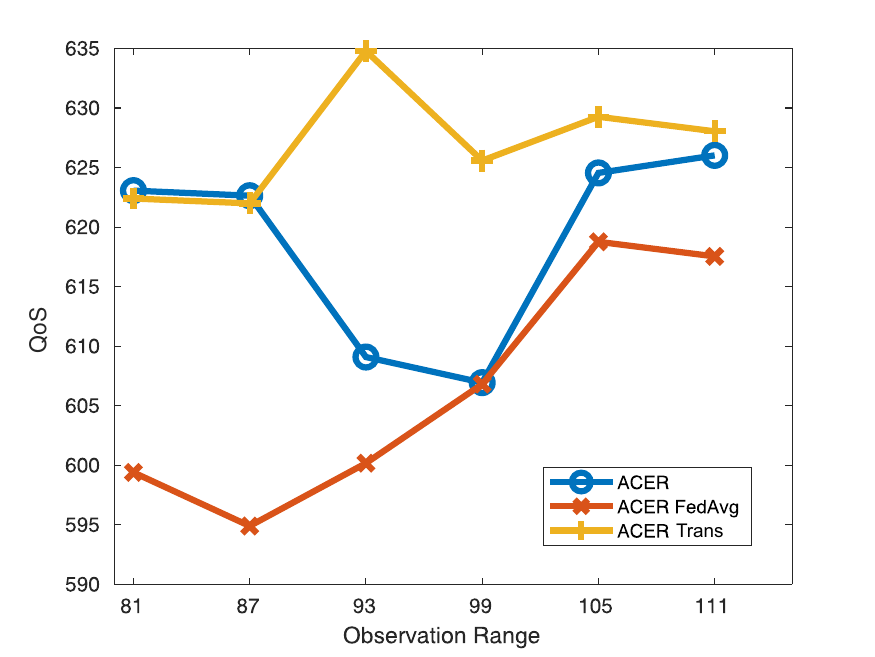}
    \caption{The QoS performance difference in observation ranges between ACER, ACER FedAvg, and ACER FedTrans.}
    \label{fig:obsrange}
    \end{minipage}
\end{figure}

\paragraph{Federated and Transfer Learning}
Following our analysis of how federated learning supports multi-agent cooperation, federated learning can share other agents' policies and stabilize the learning procedure in a non-stationary environment. As shown in Fig.\ref{fig:average_steg}, federated learning can effectively improve the QoS performance. However, the ACER FedAvg removes personal characteristics in each agent, which can cause a convergence problem if the learnt global model is not generalized enough. In Fig.\ref{fig:average_steg}, the QoS performance of ACER FedAvg drop after achieving similar QoS performance as a fixed scheme. It is worth mentioning that the policy generated by ACER FedAvg around the peak QoS performance is also quite similar to the fixed scheme shown in Fig.\ref{fig:fixcoop}. After that, the QoS performance drops due to the extremely biased experience from agents. There are two reasons for this. The fixed scheme is a locally optimal policy. As in our case, we only have QoS performance gain for cooperation, which means that the fixed scheme maximises the cooperating agents and it is a local optimal. After reaching this local optimal, the agents can usually expect a cooperation cluster to be formed by always choosing certain cooperation decisions for expected high reward. Then, the policy is set stationary to a certain action. Such knowledge is harmful and useless for the global model when FedAvg is performed. Another reason is that over half of the agents ($14$ in  $20$) are located on the edge of the network in our simulation environment, whose experience highly differs from each other. The agent on the edge can only cooperate with a limited amount of neighbours ($2$ in the corner and $4$ in the edge). Thus, performing the FedAvg method and fully accepting this incomplete and highly personalized knowledge results in a meaningless global model, and the agents can never learn to escape the local optimum with such a global model. Thus, the QoS performance drops after achieving sub-optimal solutions. It highlights the necessity of employing transfer learning approaches to share the common knowledge and maintain the personality of each agent without simply averaging and replacing. As shown in Fig.\ref{fig:average_steg}, the ACER FedTrans algorithm effectively supports the cooperation and significantly outperform the ACER FedAvg algorithm and Fixed scheme.

Then, we show the influence of federated frequency on the QoS performance, as it is critical for a federated learning system where communication costs matter.
Following our theoretical analysis in Fig.\ref{fig:convergespeed}, the federated frequency can influence the convergence speed of the algorithm, i.e. the higher federated frequency, the faster the convergence speed. But federated learning can still significantly improve the convergence speed with relatively low averaging frequency. We exam this analytical result with our simulation environment and present the result in Fig.\ref{fig:federated_steps}. At the early stage of learning ($0-2000$ epoch), the learning process with low federated averaging frequency ($F=320-1280$) converge faster the early stage of the learning. Because it allows personal characteristics in each agent and converges to local optimal shown in Fig.\ref{fig:fixcoop}, while the frequent federated operation prevents the agents to stick to this local optimal. In the later stage (after $2000$ epoch), the agent search to jump out of the local optimum which requires information from other agents to further improve the QoS performance. Thus, the high federated averaging frequency ($F=20-160$) efficiently supports the learning process with less difference in agents' policies information, which achieve slightly higher QoS performance and converge speed than the one with low federated averaging frequency ($F=320-1280$). 

% \begin{figure}[htbp!]
%     \centering
%     \includegraphics[width=0.4\textwidth]{F_Steg.pdf}
%     \caption{The influence of federated frequency on the performance of ACER algorithm with FedTrans.}
%     \label{fig:federated_steps}
% \end{figure}

% \begin{figure}[htbp!]
%     \centering
%     \includegraphics[width=0.4\textwidth]{epsnoneps.pdf}
%     \caption{The difference in learning performance with episodic and non-episodic environment.}
%     \label{fig:eps_noeps}
% \end{figure}

% \paragraph{Environment Settings}
% In Fig.\ref{fig:eps_noeps}, we also show the performance difference on our considered algorithms with episodic ($T<\infty$) and non-episodic ($T=\infty$) environment. As communication scenarios are mostly non-episodic, it is necessary to take this important factor when designing algorithms. Our results show that there will be a great difference in performance with the episodic and non-episodic environment. With our architecture, the ACER-Q algorithm can achieve the best performance and fast convergence speed with an episodic environment, while its performance in non-episodic learning is much worse. Meanwhile, the ACER algorithm with FedTrans converges in non-episodic learning. There is limited analysis for the reason of this phenomenon \cite{Naik2019Discounted}, which can be an important topic in further study. 

In Fig.\ref{fig:distribution}, we investigate the influence of user distribution for our federated algorithms. We assume the users are distributed in PCP instead of PPP, as PCP is realistic in modelling the users' positions. PPP is easier for the multi-agent algorithm to learn since the users' positions around each agent are homogeneous, which is not the case for PCP. The PCP introduces challenges for cooperative multi-agent algorithms. As shown in Fig.\ref{fig:distribution}, the ACER with PPP distributed users can achieve the highest QoS performance even without the federated algorithm, which is always the fixed scheme shown in Fig.\ref{fig:fixcoop}. However, the ACER performs the worse with PCP distributed users, which requires certain common knowledge between agents. The algorithms with the support of the federated algorithms (ACER FedTrans) can obtain good QoS performance with PCP distributed users. This shows the potential issues and future directions for current cognitive network simulations.

In Fig.\ref{fig:obsrange}, we study the influence of the observation range for our algorithms. We pick six different observation range: $81$ ($39\%$ coverage of neighbourhood's effective area), $87$ ($44\%$), $93$ ($49\%$), $99$ ($54\%$), $105$ ($60\%$), $111$ ($65\%$). The QoS performance of ACER FedAvg increases with the increasing knowledge of the neighbour's effective area, which matches our analysis. ACER fails to converge without enough knowledge from the neighbour's effective area. The ACER Trans performs well for all different observation range by sharing the knowledge without violating the personalities of each agent. It is worth mentioning that ACER and ACER Trans algorithm with $81$ observation range actually generates the fixed policy in Fig.\ref{fig:fixcoop} with little information from neighbours.

\section{Conclusion}
In this paper, we introduced a federated multi-agent RL architecture to solve the scalability problem in communication scenarios by decomposing the optimization function geometrically. We highlighted that federated learning can effectively accelerate the convergence speed and enhance cooperation. We investigated the theoretical basis for the benefit of federated learning. We also derived an upper-bound of federated multi-agent system with different federated frequency. We have shown the necessity of using transfer learning to transfer knowledge from global model to local model with the existence of centralized-decentralized mismatch. We then examined our result with a coordinated multi-point scenario. Our results demonstrated that our architecture can effectively handle the cooperation scenario with the relatively large amount of participating APs. We have also shown that the transfer learning methods outperform the federated averaging algorithm, which matches our analysis. The simulation results have shown that the large observation range can help the cooperation. These findings provide design insights for the future development of multi-agent algorithms in wireless communication networks.


\end{document}